\documentclass[a4paper,11pt]{article}
\pdfoutput=1 
\usepackage{jheppub}
\usepackage[T1]{fontenc} 
\usepackage{mathtools}
\usepackage{color}
\usepackage{hyperref}
\usepackage{subcaption}
\usepackage{slashed}
\usepackage{cancel}
\usepackage{bm}
\usepackage{array}
\usepackage{upgreek}
\usepackage[normalem]{ulem}
\makeatletter

\definecolor{SCGPblue}{rgb}{0.2196, 0.3255, 0.6431}

\hypersetup{colorlinks=true, urlcolor=SCGPblue,linkcolor=SCGPblue,citecolor=SCGPblue}

\let\originalleft\left
\let\originalright\right
\renewcommand{\left}{\mathopen{}\mathclose\bgroup\originalleft}
\renewcommand{\right}{\aftergroup\egroup\originalright}

\makeatletter
\gdef\@fpheader{\phantom{YITP}}
\makeatother

\title{Baryon Junction and String Interactions: Part II}

\author[1]{Xuzixiang Lou}
\author[1,2]{and Siwei Zhong}

\affiliation[1]{Simons Center for Geometry and Physics, Stony Brook University,\\Stony Brook, NY 11794, USA}
\affiliation[2]{C. N. Yang Institute for Theoretical Physics, Stony Brook University,\\Stony Brook, NY 11794, USA}

\emailAdd{xuzixiang.lou@stonybrook.edu}
\emailAdd{siwei.zhong@stonybrook.edu}

\abstract{We study junctions between confining strings, extending the analysis of \cite{Komargodski:2024swh}. These junctions arise in Yang-Mills theories, and we focus on their universal low-energy dynamics. Using open-closed duality, we map junctions with nonlinear corrections to the $s$-wave scattering amplitudes between confining string loops. In $(3+1)$ dimensions, we uncover an accidental $\mathbb{Z}_2$ symmetry. This symmetry implies novel selection rules for loop scattering amplitudes and is broken by the junction mass at subleading order. We determine the total mass of baryons up to order $(\text{baryon size})^{-3}$, providing new, testable predictions for lattice simulations.}

\begin{document} 
\maketitle
\flushbottom

\section{Introduction and summary}

Confinement is a fascinating phenomenon at the heart of modern physics. Since the discovery of asymptotic freedom \cite{Gross:1973id, Politzer:1973fx}, sustained efforts have been devoted to understanding the mechanism in the Standard Model that confines color quarks inside hadrons. Evidence from lattice simulations \cite{Bali:1994de, Bali:1997am} and simplified theoretical models \cite{Nielsen:1973cs, Mandelstam:1974pi} indicates that the chromoelectric flux between quarks forms string-like excitations with nonzero tension, known as the QCD flux tubes. This phenomenon also occurs in type-II superconductors, where magnetic monopoles are confined by the Abrikosov vortices \cite{Abrikosov:1956sx}. Other models exhibiting similar confining strings include $(1+1)$-dimensional QED \cite{Schwinger:1962tp}, $(2+1)$-dimensional $U(1)$ gauge theory with monopole-instantons \cite{Polyakov:1976fu}, $(2+1)$-dimensional $\mathbb{Z}_2$ gauge theory \cite{Wegner:1971app,Fradkin:1978dv}, and Seiberg--Witten theory with softly broken supersymmetry \cite{Seiberg:1994rs}. When local excitations are fully gapped, these string-like excitations are stable against breaking and can only end on probe particles or other strings.

The dynamics of the confining strings are elusive at first sight, since the underlying field theory is often strongly coupled at long distances (e.g., $(3+1)$-dimensional QCD). For sufficiently long strings, however, it was shown under mild assumptions that their low-energy dynamics are uniquely determined by the Nambu--Goto action \cite{Luscher:2004ib, Aharony:2009gg, Dubovsky:2012sh, Aharony:2013ipa}. This action describes the shape fluctuations of a string with tension $l_s^{-2}$ and is given by \cite{doi:10.1142/9789812795823_0026, Goto:1971ce}
\begin{equation}
\label{eq_NG action 1}
S_\text{NG}=-l_s^{-2}\int d t d\sigma \sqrt{-\det \left(\partial_\upalpha X^\mu \partial_\upbeta X_\mu \right)}~,
\end{equation}
where $X^\mu(t,\sigma)$ is the embedding of the string worldsheet in $(d+1)$-dimensional spacetime. For short strings, quantization of the Nambu–Goto action spoils Lorentz invariance outside the critical dimension $d=25$. Nevertheless, the action \eqref{eq_NG action 1} defines a consistent effective theory for fluctuations around long-string backgrounds in arbitrary dimension $d$. It thereby follows that long confining strings in different theories share common properties that are controlled by the string tension $l_s^{-2}$ at low energies.

The junctions are important structures that connect multiple confining strings. They are directly observed inside baryonic states in lattice simulations of $(3+1)$-dimensional QCD \cite{Takahashi:2000te, Takahashi:2002bw, Bissey:2005sk, Bissey:2006bz}, and are commonly referred to as baryon junctions. When the strings are associated with $1$-form symmetries, these junctions exist provided they satisfy the fusion rules. For example, multiple Nielsen--Olesen strings can meet at a junction when their magnetic charges sum to zero. For a partial list of early and recent studies on the baryon junctions, see, e.g., \cite{ARTRU1975442, Rossi:1977cy, Kharzeev:1996sq, Witten:1998xy, Jahn:2003uz, Imamura:2004tf, Pfeuffer:2008mz, Bakry:2014gea, Altmann:2024odn, Komargodski:2024swh}.

Do baryon junctions also exhibit universal low-energy dynamics? In our previous work \cite{Komargodski:2024swh} and this paper, we address this question in the effective field theory framework. We focus on trivalent junctions and argue that their low-energy dynamics are controlled by the junction mass $M$. In particular, we determine the ground state energy of a probe baryon, consisting of three identical strings of length $L$ meeting at a junction, up to  order $L^{-3}$:
\begin{equation}
\mathtoolsset{multlined-width=0.9\displaywidth}
\begin{multlined}
E_{\text{GS}}=\frac{3L}{l_s^2}+M-\frac{(d-2)\pi}{16L}-\frac{ (d+2)\pi  M l_s^2}{144 L^2}\hfill\\
\hfill\boxed{{}+\frac{(d+6)\pi M^2l_s^2}{432L^3}-\frac{(d-2)^2\pi^2l_s^2}{1536L^3}}+O\left(L^{-4}\right)~,
\end{multlined}
\end{equation}
where the boxed $L^{-3}$ term provides a new prediction for lattice simulations. We note that the expansion coefficients, up to high orders, are fixed by the junction mass $M$ and the string tension $l_s^{-2}$. 

The simple low-energy dynamics of the baryon junction have deep implications for the interactions between long closed strings (i.e., glueball-like excitations). These two seemingly different subjects are related by the \textit{open-closed duality} \cite{Polchinski:1987tu, Cardy:1989ir}, which we analyze explicitly in this work. Crucially, we map the baryon junction to local interactions that govern the $s$-wave scattering of 
confining string loops. For strings wrapped on a circle of circumference $2\pi R$, we find that the scattering amplitudes pick up the overall factor
\begin{equation}
    e^{-2\pi R M}~.
\end{equation}
Consequently, interactions between large loops are exponentially suppressed when the junction mass is positive. From consistency conditions, we explicitly determine the interaction vertices at leading orders in the $R^{-1}$ expansion.

From short to long distances, a natural question is what physical scale sets the junction mass $M$ in the effective field theory. We generally expect that $M\sim l_s^{-1}$ and is set by the mass gap of the underlying quantum field theory. In the large $N$ expansion, $M$ scales as the total energy of attached strings\footnote{In the ’t Hooft limit, the junction mass is of order $N$ for D5-brane realizations and order $N^2$ for NS5-brane realizations.}, and the junction becomes parametrically heavy \cite{Witten:1979kh, Witten:1998xy, Imamura:2004tf}. In this paper, we identify a novel $\mathbb{Z}_2$ symmetry for strings in $(3+1)$-dimensional spacetime that is broken by the junction mass $M$ at subleading order. See equation \eqref{eq_duality defect}. It would be intriguing to understand whether this $\mathbb{Z}_2$ symmetry can be realized at short distances and thereby requires $M=0$.

The rest of the paper is organized as follows. Section \ref{sec_review} provides a brief review of effective string theory. In Section \ref{sec_The baryon junction condition}, we summarize the effective description of the baryon junction introduced in \cite{Komargodski:2024swh}. We establish the open-closed duality of the baryon junction in Section \ref{sec_Open-closed duality of baryon junctions}. In particular, the worldsheet formulation of the novel $\mathbb{Z}_2$ symmetry and its implications in the closed channel are discussed in Section \ref{sec_symmetry constraints}. Section \ref{sec_Higher-loop corrections} extends the open-closed duality to higher orders, incorporating nonlinear corrections from the confining strings and the baryon junction. Finally, we discuss quantum corrections to the ground state energies (i.e., masses) of probe baryons in Section \ref{sec_ground state energies of probe baryons} and Appendix \ref{sec_app_rational points}.

We close by commenting on the relation to our previous work. In \cite{Komargodski:2024swh}, the open-closed duality of baryon junctions was analyzed for a special string-junction configuration with linear quantum fluctuations. In this paper, we extend the picture to general configurations as in figure \ref{pic_baryon config} and also include higher-order nonlinear corrections. Some results from \cite{Komargodski:2024swh} are reproduced and generalized in equations \eqref{eq_coupling const 1}, \eqref{eq_coupling const 2} and \eqref{eq_gs energy to order 2}.

\section{Review of the effective string theory}
\label{sec_review}

We first review the effective field theory of long confining strings in $(d+1)$-dimensional spacetime. See also \cite{Polchinski:1991ax,Luscher:2004ib,Drummond:2004yp,Meyer:2006qx,Aharony:2009gg,Aharony:2010db,Aharony:2010cx,Aharony:2011gb,Dubovsky:2012sh,Meineri:2012uav,Aharony:2013ipa,Brambilla:2014eaa,Hellerman:2014cba,Brandt:2016xsp,Hellerman:2017upi,EliasMiro:2019kyf,EliasMiro:2021nul,Caselle:2021eir, Komargodski:2024swh, Albert:2026fqj} for a necessarily incomplete selection of references. 

From a modern perspective, an infinitely long confining string is a vacuum of the underlying UV theory that spontaneously breaks the Poincar\'e symmetry $ISO(d,1)$ down to $ISO(1,1)\times SO(d-1)$. The group $ISO(1,1)$ denotes the residual Poincaré symmetry along the string direction, whereas $SO(d-1)$ is the transverse rotation group. This symmetry-breaking pattern gives rise to exactly $(d-1)$ Nambu--Goldstone Bosons (NGBs) \cite{Low:2001bw} that describe transverse fluctuations of the string. In the rest of this paper, we assume that there are no other degrees of freedom in the low-energy effective theory of the string other than these NGBs.\footnote{In supersymmetric gauge theories, there are also fermionic modes localized on the BPS confining strings \cite{Auzzi:2003fs, Hanany:2004ea, Edalati:2007vk}. These cases are beyond the scope of this paper.} This assumption applies to a wide class of gapped confining gauge theories, including the pure $SU(N)$ Yang-Mills theory in $(3+1)$ and $(2+1)$ dimensions \cite{Athenodorou:2010cs, Athenodorou:2011rx, Caselle:2011fy}.

The effective action of NGBs localized on the confining string is strongly constrained by the nonlinearly realized Poincar\'e symmetry and by worldsheet diffeomorphism invariance. The leading terms in the low-energy expansion of the effective string theory are dominated by the Nambu--Goto action \eqref{eq_NG action 1}, such that
\begin{equation}
\label{eq_NG action 2}
S_\text{string}=-\int d t d\sigma \sqrt{-\det \left(\partial_\upalpha X^\mu \partial_\upbeta X_\mu \right)}\left[l_s^{-2}+O(\partial^6)\right]~.
\end{equation}
The leading higher-derivative corrections in the action \eqref{eq_NG action 2} include the Polyakov–Kleinert rigidity term \cite{Polyakov:1986cs, Kleinert:1986bk} at $O(\partial^6)$. Taking advantage of the diffeomorphism invariance, we can choose the static gauge of the worldsheet embedding $X_\mu(t,\sigma)$ as
\begin{equation}
\label{eq_static gauge}
    X_0=t~,~~X_1=\sigma~,~~\text{and}~~X_i=l_s x_i(t,\sigma)\text{ for }2\leq i\leq d~.
\end{equation}
The effective string action \eqref{eq_NG action 2} under the ghost-free gauge \eqref{eq_static gauge} takes the expansion form
\begin{equation}
\label{eq_NG action 3}
\mathtoolsset{multlined-width=0.9\displaywidth}
\begin{multlined}
S_{\text{string}} =\int dt d\sigma \left[-\frac{1}{l_s^2}+\frac{1}{2}\left((\partial_t x_i)^2-(\partial_\sigma x_i)^2\right)\right.\hfill\\
\hfill \left.+\frac{l_s^2}{8}(\partial_t x_i-\partial_\sigma x_i)^2(\partial_t x_{{i}'}+\partial_\sigma x_{{i}'})^2+O(\partial^6)\right],
 \end{multlined}
\end{equation}
where both the $O(\partial^2)$ quadratic term and the $O(\partial^4)$ quartic term follow from the Nambu--Goto action.

We now specify our EFT order counting convention as follows: terms in the confining string worldsheet action are of order $n$ if they scale as $O(\partial^{n+2})$; terms supported on the 1-dimensional submanifold of the worldsheet are of order $n$ if they scale as $O(\partial^{n+1})$. In this convention, classical contributions to the path-integral are denoted by terms with negative orders, whereas the leading quantum corrections start at order $0$. For example, the three terms in the action \eqref{eq_NG action 3} are of orders $-2$, $0$, and $2$, respectively. The order $0$ quadratic term in \eqref{eq_NG action 3} is simply the kinetic term of free worldsheet bosons, whereas the order $2$ quartic term corresponds to a $T\Bar{T}$-deformation of the free theory \cite{Caselle:2013dra, Dubovsky:2014fma, Chen:2018keo}. Notably, the effective string action \eqref{eq_NG action 3} is completely determined by the classical string tension $l_s^{-2}$ up to order $3$.

The action \eqref{eq_NG action 3} can be used to compute the spectrum of a long, closed confining string with high precision. We consider a string that wraps along the periodic $X_1$-direction of circumference $2\pi R$, such that in the static gauge \eqref{eq_static gauge} we have $\sigma \sim \sigma+2\pi R$. The energy of a generic closed string state takes the following form \cite{Arvis:1983fp, Luscher:2004ib, Aharony:2010db, Aharony:2013ipa}
\begin{equation}
\label{eq_closed energy level}
E^{\text{closed}}_a=\frac{2\pi R}{l_s^2}+\frac{2}{R}\left(n_a-\frac{d-1}{24}\right)-\frac{l_s^2}{\pi R^3}\left(n_a-\frac{d-1}{24}\right)^2+O(R^{-5})~,
\end{equation}
where the subscript $a$ (and similarly $b$ and $c$ below) labels weakly coupled multi-particle states on the closed string. In equation \eqref{eq_closed energy level}, $n_a$ denotes the average of the left- and right-moving excitation levels, following from the order $0$ free kinetic term in the action \eqref{eq_NG action 3}\footnote{In the free theory approximation of \eqref{eq_closed energy level}, the worldsheet Hamiltonian of the closed string reads \begin{equation*}
    H_{\text{free}}=\sum_{i=2}^d\sum_{n \in \mathbb{N^+}} \frac{n}{R}\left((\alpha^i_{\text{L},n})^{\dagger} \alpha^i_{\text{L},n}+(\alpha^i_{\text{R},n})^{\dagger} \alpha^i_{\text{R},n}\right)-\frac{d-1}{12R}~,
\end{equation*}
where $\alpha^i_{\text{L},n}$ and $\alpha^i_{\text{R},n}$ are the $n$-th annihilation operator left- and right-moving oscillator modes, respectively. Eigenvalues of $H_{\text{free}}$ take the form of $\frac{2}{R}(n_a-\frac{d-1}{24})$, as explained below equation \eqref{eq_closed energy level}.}. For states with vanishing momentum along the $X_1$-direction, we have $n_a \in \mathbb{N}$.
 
 We note from \eqref{eq_closed energy level} that there exists a large degeneracy between closed string excited states at the precision of $O(R^{-3})$, which is a consequence of the Poincar\'e symmetry and the integrability of the quartic deformation in \eqref{eq_NG action 3}. In the following sections, we will be interested in states that transform in the scalar representation of the transverse rotation group $SO(d-1)$ and carry zero longitudinal momentum. The two lowest-lying states satisfying these conditions are the ground state $\textbf{0}$ (with $n_{\textbf{0}}=0$) and an excited state $\textbf{1}$ (with $n_{\textbf{1}}=1$), while states with excitation levels $n_a\geq 2$ are degenerate.

\subsection{Open string boundary conditions}
\label{sec_open string boundary conditions}

In this subsection, we review the Dirichlet and the Neumann boundary effective theories of open confining strings. See, e.g., \cite{Luscher:2004ib, Aharony:2009gg, Aharony:2010cx} for a systematic and detailed analysis.

Let us consider a semi-infinite string with longitudinal coordinate $X_1=\sigma\geq 0$, whose worldsheet action is given by \eqref{eq_NG action 3}. The variational principle of the order $0$ kinetic term in \eqref{eq_NG action 3} leads to two canonical choices of boundary conditions at $\sigma=0$, namely Dirichlet and Neumann. The Dirichlet boundary condition that preserves $SO(d-1)$ rotation symmetry reads
\begin{equation}
\label{eq_Dirichlet bdry condition}
    \text{Dirichlet}~:~~~\partial_tx_i=0~,~~\text{for}~~2\leq i \leq  d~.
\end{equation}
It is also convenient to set the NGBs $x_i=0$ at the Dirichlet boundary $\sigma=0$. Higher-order corrections of the Dirichlet boundary are constrained by Poincar\'e symmetry, and the leading boundary deformation takes the following form:
\begin{equation}
\label{eq_Dirichlet action}
    S_{\text{Dirichlet}}=\int_{\sigma=0} dt \left[-m_\text{D}+\kappa_\text{D}(\partial_t\partial_\sigma x_i)^2+O\left(\partial^6\right)\right]~.
\end{equation}
The $m_\text{D}$ term in \eqref{eq_Dirichlet action} is of order $-1$ according to our EFT counting convention, whereas the $\kappa_\text{D}$ term is of order $3$. The constant $m_\text{D}$ simply shifts the zero-point energy associated with the Dirichlet boundary. In lattice simulations, $m_\text{D}$ represents the scheme-dependent mass of the static quarks. For convenience, we set $m_\text{D}=0$ throughout this paper, i.e., we subtract the quark mass contributions.

On the other hand, the $SO(d-1)$ symmetric Neumann boundary condition reads
\begin{equation}
    \text{Neumann}~:~~~\partial_\sigma x_i=0~,~~\text{for}~~2\leq i \leq  d~.
\end{equation}
Similarly, we can consider higher-order corrections to the Neumann boundary condition that are compatible with the Poincar\'e symmetry. The leading deformations are as follows 
\begin{equation}
\label{eq_Neumann action}
\begin{aligned}
    S_{\text{Neumann}}=&\int_{\sigma=0} dt \left[-m_{\text{N}}\sqrt{-(\partial_tX_\mu)^2}+\kappa_\text{N}(\partial_t^2x_i)^2+O\left(\partial^6\right)\right]\\
    =&\int_{\sigma=0} dt \left[-m_{\text{N}}+\frac{m_{\text{N}}l_s^2}{2}(\partial_t x_i)^2-\frac{m_{\text{N}}l_s^4}{8}(\partial_t x_i)^4+\kappa_\text{N}(\partial_t^2x_i)^2+O\left(\partial^6\right)\right]~.
\end{aligned}
\end{equation}
The first term in \eqref{eq_Neumann action} denotes the worldline length of the dynamical string endpoint, where the parameter $m_{\text{N}}$ is interpreted as the endpoint mass. We note that both the order $-1$ and order $1$ terms in \eqref{eq_Neumann action} are fixed by the mass $m_{\text{N}}$, whereas at order $3$ an additional parameter $\kappa_\text{N}$ can be introduced. Unlike $m_\text{D}$ in \eqref{eq_Dirichlet action}, $m_{\text{N}}$ perturbs the spectrum associated with the Neumann boundary.

\subsection{Open-closed duality}
\label{sec_The open-closed duality}

We now follow \cite{Luscher:2004ib, Meyer:2006qx, Giudice:2009di, Aharony:2010cx} and discuss the duality that associates the open string spectrum with an effective theory of closed string fields. For simplicity, we consider a finite string extending from $X_1=\sigma=0$ to $X_1=\sigma=L$, with the Dirichlet boundary condition \eqref{eq_Dirichlet bdry condition} at both ends. This confining string configuration describes a probe meson.

The partition function of the finite open string, denoted by $\mathcal{Z}$, can be computed and interpreted in several equivalent ways. In the open channel, the partition function is given by the Boltzmann sum
\begin{equation}
\label{eq_2pt open channel}
   \text{open channel}~:~~~ \mathcal{Z}=\sum_{E_{\text{open}}}e^{-\beta E_{\text{open}}}~,
\end{equation}
where $E_{\text{open}}$ denotes the energy eigenvalues of open string states defined on $0\leq \sigma \leq L$, and $\beta$ is the inverse temperature. The partition function can be obtained by Wick rotating $t\to -i \tau$ and compactifying $\tau \sim \tau+\beta$ in the path-integral of the effective string action \eqref{eq_NG action 3}. In particular, we take the circumference of the Euclidean time circle to be $\beta=2\pi R$. The open string partition function takes the following form
\begin{equation}
\label{eq_meson partition function 1}
\mathcal{Z}=\frac{e^{-2\pi R L/l_s^2}}{\left[\eta(q)\right]^{(d-1)}}\left[1-\frac{(d-1)\pi l_s^2 }{1152L^2}\ln q\left(2E_4(q)+(d-3)(E_{2}(q))^2\right)+O\left(\partial^3\right)\right]~,
\end{equation}
where the modular parameter $q\equiv \exp (-2\pi^2 R/L)$, $\eta(q)$ denotes the Dedekind eta function, and $E_2(q)$, $E_4(q)$ are the Eisenstein series (see Appendix \ref{sec_app_modular function}). In the expansion \eqref{eq_meson partition function 1}, terms that scale as $O(L^{-n_1}R^{-n_2})$ have been collectively denoted by $O(\partial^{n_1+n_2})$. We note that order $n$ terms in the effective action give leading contributions to the partition function that scale as $O(\partial^n)$. For example, corrections to the string endpoints, parametrized by $\kappa_{\text{D}}$ \eqref{eq_Dirichlet action} at $\sigma=0$ and $\sigma=L$, are grouped into $O(\partial^3)$ terms in the expansion \eqref{eq_meson partition function 1}.

In the closed channel, the partition function is interpreted as the two-point function of the Polyakov loop operators wrapped along the $X_0$-direction. A wrapped Polyakov loop preserves the Lorentz symmetry in the $d$-dimensional space spanned by the coordinates $\vec{X}=(X_1, X_{2\leq i\leq d})$, and it defines a scalar point operator $\Omega (\vec{X})$. In general, the reducible operator $\Omega$ admits a decomposition into massive particle fields in the gapped effective theory as follows: 
\begin{equation}
\label{eq_Polyakov decomposition}
    \Omega(\vec{X})= (\pi l_s^2)^{\frac{d-1}{4}}\sum_{a} v_a\sqrt{E^{\text{closed}}_a} \Phi_a(\vec{X})~,
\end{equation}
where $v_a \in \mathbb{C}$ denote decomposition coefficients, and $\Phi_a$ are $d$-dimensional complex scalar fields with definite masses. In equation \eqref{eq_Polyakov decomposition} (and similarly below), the summation runs over Dirichlet boundary states of a confining closed string with circumference $2\pi R$. These boundary states reduce to Ishibashi states in the free theory limit, and are perturbed by higher-order corrections in the action \eqref{eq_NG action 3} and \eqref{eq_Dirichlet action}. These closed string states satisfy the following properties: (i) they carry zero longitudinal momentum, as required by the Cardy condition \cite{Cardy:1989ir}; and (ii) they transform in the scalar representation of the transverse rotation group $SO(d-1)$, which is preserved by the Dirichlet condition \eqref{eq_Dirichlet bdry condition}. As we noted previously, the two lowest-lying boundary states are the ground state $\textbf{0}$ and the excited state $\textbf{1}$. We have also chosen the normalization in the decomposition \eqref{eq_Polyakov decomposition} such that the coefficients $v_a$ are dimensionless.

We adopt the following effective action ansatz for the dynamics of the closed-string fields $\Phi_a$ in $d$-dimensional space:
\begin{equation}
\label{eq_closed string EFT}
    S_{\text{loops}}=\int d^d\vec{X}\left[\frac{1}{2}\sum_{a}\left(|\partial_\nu\Phi_a|^2+(E^{\text{closed}}_a)^2|\Phi_a|^2\right)+V(\Phi_a,\Phi_b,...)\right]~,
\end{equation}
where $V(\Phi_a,\Phi_b,...)$ denotes interaction terms. The functional $V(\Phi_a,\Phi_b,...)$ is constrained by 1-form symmetries of the confining strings, whether IR emergent or UV exact. We will investigate the structure of $V(\Phi_a,\Phi_b,...)$ in Sections \ref{sec_Open-closed duality of baryon junctions} and \ref{sec_Higher-loop corrections}. Notably, the mass of the field $\Phi_a$ corresponding to the closed-string state $a$ is given by \eqref{eq_closed energy level}. From a $d$-dimensional viewpoint, $E^{\text{closed}}_a$ plays the role of the internal mass of a particle at rest. 

Following the decomposition \eqref{eq_Polyakov decomposition} and the effective action \eqref{eq_closed string EFT}, we find that the closed channel partition function takes the form
\begin{equation}
\label{eq_2pt closed channel}
   \text{closed channel}~:~~~ \mathcal{Z}=\langle \Omega^\dagger(\vec{X})\Omega(\vec{{X}'})\rangle= \sum_{a}|v_a|^2\frac{(E^{\text{closed}}_a)^{\frac{d}{2}}l_s^{d-1}}{\sqrt{\pi}(2L)^{\frac{d-2}{2}}}K_{\frac{d-2}{2}}\left(E^{\text{closed}}_a L\right)~,
\end{equation}
where $|\vec{X}-\vec{{X}'}|=L$, and $K_{\alpha}(x)$ denotes the modified Bessel function of the second type. In equation \eqref{eq_2pt closed channel}, we have used the tree-level propagator of massive particles in $d$-dimensional space and omitted possible loop corrections from the interaction terms. Such loop corrections are exponentially suppressed by the mass scale $O(E^{\text{closed}}_a)=O(R/l_s^2)$, and the partition function is therefore heavily dominated by the contribution in \eqref{eq_2pt closed channel}. With the dual modular parameter given by $\tilde{q}\equiv \exp(-2 L/R)$, the partition function \eqref{eq_2pt closed channel} admits the expansion
\begin{equation}
\mathtoolsset{multlined-width=0.9\displaywidth}
\begin{multlined}
\label{eq_meson partition function 2} 
\mathcal{Z}=\left(\frac{\pi R}{L}\right)^{\frac{d-1}{2}}\frac{e^{-2\pi R L/l_s^2}}{\tilde{q}^{\frac{d-1}{24}}}\sum_a|v_a|^2\tilde{q}^{n_a}\left[1-\frac{(d-1)l_s^2}{2\pi R^2}\left(\frac{\ln \tilde{q}}{d-1}\left(n_a-\frac{d-1}{24}\right)^2\right.\right.\\
\left.-\left(n_a-\frac{d-1}{24}\right)+\frac{d-3}{4\ln \tilde{q}}\Biggr)+O\left(\partial^3\right)\right]~.
\end{multlined}
\end{equation}

The open-closed duality identifies the partition functions evaluated in the open channel \eqref{eq_2pt open channel} and in the closed channel \eqref{eq_2pt closed channel}. Consequently, each term in the expansion \eqref{eq_meson partition function 1} is matched to the corresponding term in \eqref{eq_meson partition function 2} by a modular transformation. This leads to strong constraints on the decomposition coefficients $v_a$. By locality, the coefficients $v_a=v_a(l_s,R,\kappa_\text{D}, \dots)$ are fixed solely by the local parameters of the Dirichlet boundaries, which are implemented by the Polyakov loops. From \eqref{eq_2pt open channel} and \eqref{eq_meson partition function 2}, we find that
\begin{equation}
\label{eq_v0 and v1}
\begin{aligned}
v_{\textbf{0}}={}&1+\frac{(d-1) l_s^2}{48 \pi  R^2}+O\left(R^{-3}\right)~,\\
v_{\textbf{1}}={}&\sqrt{d-1}\left(1+\frac{(d-25) l_s^2}{48 \pi  R^2}+O\left(R^{-3}\right)\right)~,
    \end{aligned}
\end{equation}
to which we will return in Sections \ref{sec_Open-closed duality of baryon junctions} and \ref{sec_Higher-loop corrections}.

\subsection{Baryon junction condition}
\label{sec_The baryon junction condition}

\begin{figure}[thb]
\centering
\includegraphics[width=.45\textwidth]{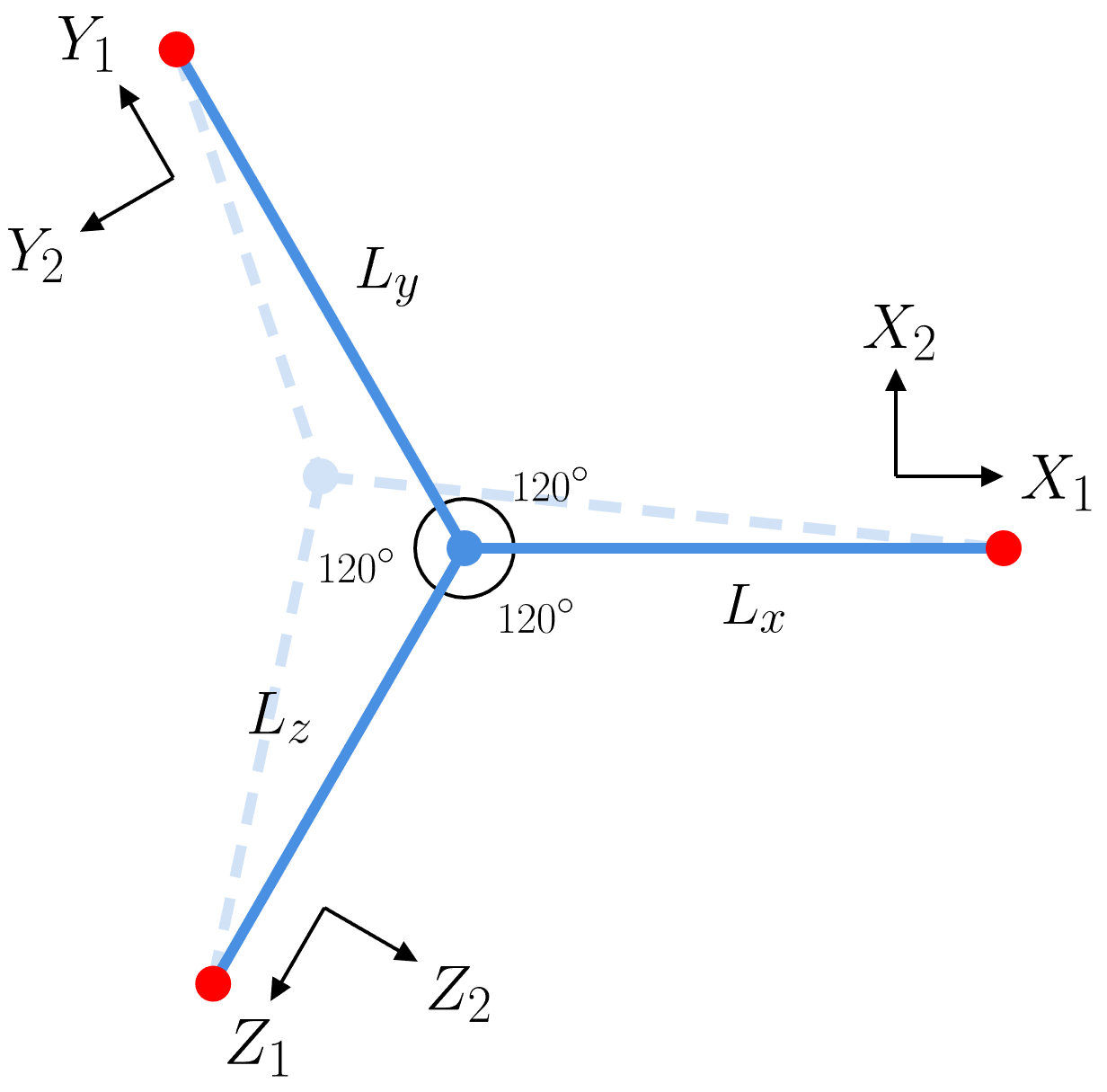}
  \caption{Confining strings in a probe baryon. The figure shows the spatial plane determined by the insertions of three static quarks (red points). The confining strings (blue lines) are tied at the dynamical baryon junction (blue point) and terminate on the quarks at their other ends.} \label{pic_baryon config}
\end{figure}

In this subsection, we review the effective field theory of the trivalent baryon junctions introduced in \cite{Komargodski:2024swh}. We consider the confining string configurations as in figure \ref{pic_baryon config}, which describe probe baryons composed of three identical quarks. Our analysis applies, for example, to baryonic states in $SU(3N)$ pure Yang-Mills theories in $(2+1)$ and $(3+1)$ dimensions. 

Classically, the junction position in the $ d$-dimensional space is determined by balancing the tensions $l_s^{-2}$ of the three confining strings. It therefore coincides with the Fermat–Weber point of the triangle whose vertices are set by the quarks. Throughout this paper, we focus on baryon configurations where the Fermat–Weber point lies within the interior of the triangle. We leave the cases where the baryon junction overlaps with one of the quarks to future work.

We denote the coordinates of three strings by $X_\mu$, $Y_\mu$, and $Z_\mu$, with their lengths $L_x$, $L_y$, and $L_z$. See also figure \ref{pic_baryon config}. In the static gauge \eqref{eq_static gauge}, we take $X_1=Y_1=Z_1=\sigma=0$ to be the joint baryon junction where three strings are connected, while $X_1=\sigma=L_x$, $Y_1=\sigma=L_y$, and $Z_1=\sigma=L_z$ are the string endpoints on the quarks. Furthermore, we use $x_i$, $y_i$, and $z_i$ for the NGBs on the corresponding worldsheet. In our previous work \cite{Komargodski:2024swh}, we argued that the effective action of the probe baryon takes the following form:
\begin{equation}
\mathtoolsset{multlined-width=0.9\displaywidth}
\begin{multlined}
\label{eq_baryon action}
S_{\text{baryon}}=\left(S_\text{strings}^{(-2)}+S_\text{strings}^{(0)}+S_\text{strings}^{(2)}\right)+\left(S_\text{junction}^{(-1)}+S_\text{junction}^{(1)}\right)\\
+\left(S_\text{displacement}^{(1)}+S_\text{displacement}^{(2)}\right)+O(\partial^3)~,
\end{multlined}
\end{equation}
which we now unpack term by term.

In equation \eqref{eq_baryon action}, the terms $S_\text{strings}^{(-2)}$, $S_\text{strings}^{(0)}$, and $S_\text{strings}^{(2)}$ collect effective actions of the three strings up to order $2$. As in the single string case \eqref{eq_NG action 3}, the order $-2$ classical term of three strings is as follows:
\begin{equation}
\label{eq_baryon strings order -2}
S_\text{strings}^{(-2)}=-l_s^{-2}\int d t \left (\int_0^{L_x}d \sigma+\int_0^{L_y}d \sigma+\int_0^{L_z}d \sigma \right)~.
\end{equation}
The order $0$ term $S_\text{strings}^{(0)}$ is the quadratic kinetic action that governs the leading quantum fluctuations on the worldsheets:
\begin{equation}
\label{eq_baryon strings order 0}
S_\text{strings}^{(0)}=\frac{1}{2}\int d t \int_0^{L_x}d \sigma \left[(\partial_t x_i)^2-(\partial_\sigma x_i)^2\right]+(x\to y)+(x\to z)~.
\end{equation}
At the quark endpoints, the NGBs $x_i$, $y_i$, and $z_i$ are subject to the Dirichlet boundary condition \eqref{eq_Dirichlet bdry condition}. On the other hand, the rigid geometry of confining strings at the baryon junction leads to the constraints:
\begin{equation}
\label{eq_junction conditions 1}
x_2+y_2+z_2=0~,~~\text{and}~~x_j=y_j=z_j~~\text{for}~~3\leq j \leq d~.
\end{equation}
It is convenient to introduce the linear combinations $\xi_i^{[1]}=x_i+y_i+z_i$, $\xi_i^{[2]}=x_i-y_i$, and $\xi_i^{[3]}=x_i+y_i-2z_i$ for $2\leq i \leq d$. In these new variables, the geometry constraints \eqref{eq_junction conditions 1} and the variational principle of \eqref{eq_baryon strings order 0} determine the following boundary condition at $\sigma=0$:
\begin{equation}
\label{eq_junction conditions 2}
\partial_t\xi_{2}^{[1]}=\partial_t\xi_{j}^{[2]}=\partial_t\xi_{j}^{[3]}=0~, ~~\text{and}~~\partial_\sigma\xi_2^{[2]}= \partial_\sigma\xi_2^{[3]}=\partial_\sigma\xi^{[1]}_j=0~~\text{for}~~3\leq j \leq d~.
\end{equation}
Higher-order corrections to the baryon junction boundary condition \eqref{eq_junction conditions 2} arise from interaction terms localized at $\sigma=0$. As we will show in Section \ref{sec_Higher-loop corrections}, these interactions are strongly constrained by Poincar\'e symmetry and open-closed duality. Finally, the order $2$ term $S_\text{strings}^{(2)}$ contains the quartic interactions on the worldsheets:
\begin{equation}
\label{eq_baryon strings order 2}
S_\text{strings}^{(2)}=\frac{l_s^2}{8}\int d t \int_0^{L_x}d \sigma (\partial_t x_i-\partial_\sigma x_i)^2(\partial_t x_{{i}'}+\partial_\sigma x_{{i}'})^2+(x\to y)+(x\to z)~,
\end{equation}
which follows from the Nambu--Goto action, as in equation \eqref{eq_NG action 3}.

The terms $S_\text{junction}^{(-1)}$ and $S_\text{junction}^{(1)}$ in equation \eqref{eq_baryon action} describe the worldline dynamics of the baryon junction. Let $W_\mu$ denote the spacetime coordinates of the baryon junction point. The leading worldline action compatible with Poincar\'e symmetry then takes the form:
\begin{equation}
\label{eq_baryon junction mass}
    S_\text{junction}=-M\int_{\sigma=0} dt\sqrt{-(\partial_t W_\mu)^2}= S_\text{junction}^{(-1)}+ S_\text{junction}^{(1)}+O(\partial^3)~.
\end{equation}
We have introduced a new EFT parameter $M$ in the action \eqref{eq_baryon junction mass}, which is interpreted as the classical mass of the baryon junction. Expanding the worldline action \eqref{eq_baryon junction mass} in the NGBs $x_i$, $y_i$, and $z_i$, we find the order $-1$ term
\begin{equation}
\label{eq_baryon junction order -1}
    S_\text{junction}^{(-1)}=-M\int_{\sigma=0} dt~,
\end{equation}
together with the order $1$ term
\begin{equation}
\label{eq_baryon junction order 1}
\mathtoolsset{multlined-width=0.9\displaywidth}
\begin{multlined}
S_\text{junction}^{(1)}=\frac{Ml_s^2}{18}\int_{\sigma=0} dt\left[(\partial_tx_j+\partial_ty_j+\partial_tz_j)^2\right.\hfill\\
\hfill\left.+(\partial_tx_2+\partial_ty_2-2\partial_tz_2)^2+3(\partial_tx_2-\partial_ty_2)^2\right]~.
\end{multlined}
\end{equation}
In equation \eqref{eq_baryon junction order 1}, we have chosen the linear combinations of $x_i$, $y_i$, and $z_i$ that survive the geometry constraint \eqref{eq_junction conditions 1}.

Unlike the open string boundaries reviewed in Section \ref{sec_open string boundary conditions}, the baryon junction is allowed to fluctuate in the longitudinal directions of the confining strings. The longitudinal displacements of the three strings at $\sigma=0$ are determined by the geometry in figure \ref{pic_baryon config}, and are given by
\begin{equation}
\label{eq_longitudinal displacement}
    \delta X_1=\frac{l_s}{\sqrt{3}}(z_2-y_2)~,~~\delta Y_1=\frac{l_s}{\sqrt{3}}(x_2-z_2)~,~~\text{and}~~\delta Z_1=\frac{l_s}{\sqrt{3}}(y_2-x_2)~.
\end{equation}
These displacements couple to EFT operators that generate infinitesimal deformations of the worldsheet boundary, analogous to the displacement operator in defect conformal field theory \cite{McAvity:1995zd, Liendo:2012hy, Jensen:2015swa}. For example, let us consider a single confining string whose endpoint fluctuates infinitesimally in the longitudinal direction with $\sigma \geq \delta X_1(t)$. Expanding the worldsheet action \eqref{eq_NG action 3} in powers of $\delta X_1$, we identify the following effective coupling at the boundary:
\begin{equation}
\label{eq_displacement def}
    S_{\text{displacement}}=\sum_{n\in \mathbb{N}}\int_{\sigma=0} dt(-\delta X_1)^{n+1}\partial_\sigma^{n}\left(-\frac{1}{l_s^2}+\frac{1}{2}\left((\partial_t x_i)^2-(\partial_\sigma x_i)^2\right)+O(\partial^4)\right)~.
\end{equation}
In the baryon junction case, the longitudinal displacements are promoted to dynamical fields as in equation \eqref{eq_longitudinal displacement}, while the worldsheet action up to order $2$ is dictated by \eqref{eq_baryon strings order -2}, \eqref{eq_baryon strings order 0}, and \eqref{eq_baryon strings order 2}. The geometric constraint $\delta X_1+\delta Y_1+\delta Z_1=0$ ensures that the order $-1$ coupling of the longitudinal fluctuations is trivial. The leading order $1$ coupling is therefore
\begin{equation}
\label{eq_baryon displacement order 1}
\mathtoolsset{multlined-width=0.9\displaywidth}
\begin{multlined}
S_\text{displacement}^{(1)}=\frac{l_s}{2\sqrt{3}}\int_{\sigma=0} dt \left[(y_2-z_2)\left((\partial_t x_i)^2-(\partial_\sigma x_i)^2\right)\right.\hfill\\
\hfill\left. {}+(z_2-x_2)\left((\partial_t y_i)^2-(\partial_\sigma y_i)^2\right)+(x_2-y_2)\left((\partial_t z_i)^2-(\partial_\sigma z_i)^2\right)\right]~,
\end{multlined}
\end{equation}
followed by the subleading order $2$ term,
\begin{equation}
\label{eq_baryon displacement order 2}
\mathtoolsset{multlined-width=0.9\displaywidth}
\begin{multlined}
S_\text{displacement}^{(2)}=\frac{l_s^2}{12}\int_{\sigma=0} dt \left[(y_2-z_2)^2\partial_\sigma\left((\partial_t x_i)^2-(\partial_\sigma x_i)^2\right)\right.\hfill\\
\hfill\left. {}+(z_2-x_2)^2\partial_\sigma\left((\partial_t y_i)^2-(\partial_\sigma y_i)^2\right)+(x_2-y_2)^2\partial_\sigma\left((\partial_t z_i)^2-(\partial_\sigma z_i)^2\right)\right]~,
\end{multlined}
\end{equation}
For simplicity, we have combined the longitudinal contributions from the three worldsheets in equations \eqref{eq_baryon displacement order 1} and \eqref{eq_baryon displacement order 2}.

To summarize, we have classified terms in the effective action of the probe baryon up to order $2$. We find that the effective action \eqref{eq_baryon action} is determined by two EFT parameters, namely the classical string tension $l_s^{-2}$ and the classical baryon junction mass $M$. New EFT parameters may appear at higher orders; for example, the $\kappa_\text{D}$ term in \eqref{eq_Dirichlet action} at the quark endpoints is of order $3$. The quantum fluctuations of the probe baryon are fully controlled by $l_s^{-2}$ and $M$ up to order $2$, whose physical implications we will discuss in the upcoming sections.

\section{Open-closed duality of baryon junctions}
\label{sec_Open-closed duality of baryon junctions}

In this section, we establish the open-closed duality for the junctions of confining strings. To set the stage, it is useful to recall the dual channels of a finite open string reviewed in Section \ref{sec_The open-closed duality}. The open-closed duality identifies the thermal partition function of a probe meson \eqref{eq_2pt open channel} with the two-point correlation function of Polyakov loop operators \eqref{eq_2pt closed channel}. We assume that the Polyakov operators (i.e., the quark endpoints of the confining strings) can be decomposed into massive particles in $d$-dimensional space. By matching the partition function \eqref{eq_meson partition function 1} with \eqref{eq_meson partition function 2} via modular transformations, we fix the decomposition coefficients up to their leading orders as in equation \eqref{eq_v0 and v1}.

We now turn to the probe baryon case in figure \ref{pic_baryon config}. In the open channel, the baryon partition function $\mathcal{Z}$ takes the form
\begin{equation}
\label{eq_baryon open channel}
    \text{open channel}~:~~~\mathcal{Z}=\sum_{E_{\text{baryon}}} e^{-\beta E_{\text{baryon}}}~,
\end{equation}
where $E_{\text{baryon}}$ denotes the energy eigenvalues of the baryon states. As in Section \ref{sec_The open-closed duality}, the partition function can be computed from the effective action \eqref{eq_baryon action} by Wick rotating $t\to -i \tau$ and compactifying $\tau\sim \tau +\beta$. We carry out this calculation in Sections \ref{sec_Partition function of the probe baryon} and \ref{sec_Higher-loop corrections}.

With three external quarks, we argue that the closed channel of the baryon partition function is given by the three-point function of Polyakov operators:
\begin{equation}
\label{eq_baryon closed channel}
    \text{closed channel}~:~~~\mathcal{Z}=\langle\Omega(\vec{X})\Omega(\vec{Y})\Omega(\vec{Z})\rangle~.
\end{equation}
In equation \eqref{eq_baryon closed channel}, $\vec{X}$, $\vec{Y}$, and $\vec{Z}$ denote positions in $d$-dimensional space. We adopt the convention
\begin{equation}
    \vec{X}=(L_x,0,\dots)~,~~\vec{Y}=(-\frac{1}{2}L_y,\frac{\sqrt{3}}{2}L_y,\dots)~,~~\text{and}~~\vec{Z}=(-\frac{1}{2}L_z,-\frac{\sqrt{3}}{2}L_z,\dots)~,
\end{equation}
where we have omitted the $(d-2)$ perpendicular directions, and the Fermat--Weber point of the $\vec{X}\vec{Y}\vec{Z}$ triangle is at the origin. The three-point function \eqref{eq_baryon closed channel} defines a nontrivial observable when $\mathbb{Z}_3$ is a 1-form symmetry subgroup of the underlying gauge theory. Our discussion applies, for instance, to $SU(3N)$ and $PSU(3N)$ pure Yang-Mills theories in $(3+1)$ dimensions~\footnote{The $PSU(3N)$ Yang-Mills theories in $(3+1)$ dimensions have a magnetic 1-form symmetry $\mathbb{Z}_{3N} \supseteq \mathbb{Z}_3$. In these theories, confining strings end on magnetic monopoles corresponding to 't Hooft line operators.}.

In Sections \ref{sec_s-wave scattering} and \ref{sec_Higher-loop corrections}, we show that the three-point function \eqref{eq_baryon closed channel} is dominated by the tree-level $s$-wave scattering of the massive particles $\Phi_a$ described in \eqref{eq_closed string EFT}. By matching the dual channels \eqref{eq_baryon open channel} and \eqref{eq_baryon closed channel} via modular transformations, we determine the interaction vertices in the potential function $V(\Phi_a,\Phi_b, \dots)$ that governs the $s$-wave scattering. In the rest of this section, we elaborate on this calculation up to order $1$, where $1$-loop corrections on the worldsheets are concerned. With simplifications, we continue this calculation to order $2$ in Section \ref{sec_Higher-loop corrections}. 

\subsection{Partition function of the probe baryon}
\label{sec_Partition function of the probe baryon}

We begin by considering the Euclidean path-integral of the effective action \eqref{eq_baryon action}
\begin{equation}
    \mathcal{Z}=\int \mathcal{D}x_i\mathcal{D}y_i\mathcal{D}z_i e^{-S_\text{baryon}}~,
\end{equation}
with Euclidean time compactified on a circle of circumference $\beta=2\pi R$. This path-integral admits the following perturbative expansion:
\begin{equation}
\label{eq_perturbative expansion}
\mathtoolsset{multlined-width=0.9\displaywidth}
\begin{multlined}
\mathcal{Z}=e^{-S_\text{strings}^{(-2)}-S_\text{junction}^{(-1)}}\mathcal{Z}^{(0)}\biggl[1-\langle S_\text{junction}^{(1)}\rangle-\langle S_\text{strings}^{(2)}\rangle-\langle S_\text{displacement}^{(2)}\rangle\hfill\\
\hfill +\frac{1}{2}\langle(S_\text{junction}^{(1)})^2\rangle+\frac{1}{2}\langle(S_\text{displacement}^{(1)})^2\rangle+O\left(\partial^3\right)\biggr]~,
\end{multlined}
\end{equation}
where we have defined
\begin{equation}
\label{eq_Z^0 definition}
    \mathcal{Z}^{(0)}\equiv \int \mathcal{D}x_i\mathcal{D}y_i\mathcal{D}z_i e^{-S_\text{strings}^{(0)}}~,~~\text{and}~~\langle {\cdots}\rangle\equiv \frac{1}{\mathcal{Z}^{(0)}}\int \mathcal{D}x_i\mathcal{D}y_i\mathcal{D}z_i ({\cdots})e^{-S_\text{strings}^{(0)}}~.
\end{equation}
The classical contributions to the path-integral are given by $S_\text{strings}^{(-2)}$ in equation \eqref{eq_baryon strings order -2} and $S_\text{junction}^{(-1)}$ in equation \eqref{eq_baryon junction order -1}. We readily find that
\begin{equation}
\label{eq_baryon classical contribution}
    S_\text{strings}^{(-2)}=\frac{2\pi R}{l_s^2}(L_x+L_y+L_z)~,~~\text{and}~~S_\text{junction}^{(-1)}=2\pi R M~.
\end{equation}
We have organized the quantum fluctuations in the path-integral \eqref{eq_perturbative expansion} according to their EFT order, where $\mathcal{Z}^{(0)}$ is the leading order $0$ contribution. Among the higher-order corrections, we note that the expectation values of parity-odd terms vanish. For example, $\langle S_\text{displacement}^{(1)}\rangle=0$ and  $\langle S_\text{junction}^{(1)}S_\text{displacement}^{(1)}\rangle=0$. In the rest of this section, we compute $\mathcal{Z}^{(0)}$ and $\langle S_\text{junction}^{(1)}\rangle$, thereby fixing the partition function up to order $1$. We will elaborate on the order $2$ corrections $\langle S_\text{strings}^{(2)}\rangle$, $\langle S_\text{displacement}^{(2)}\rangle$, $\langle(S_\text{junction}^{(1)})^2\rangle$, and $\langle(S_\text{displacement}^{(1)})^2\rangle$ in Section \ref{sec_Higher-loop corrections}.

The order $0$ effective action \eqref{eq_baryon strings order 0} is simply the free kinetic action for the NGBs $x_i$, $y_i$, and $z_i$. The partition function $\mathcal{Z}^{(0)}$ with the baryon junction condition \eqref{eq_junction conditions 1} and \eqref{eq_junction conditions 2} can be evaluated as follows~\cite{Jahn:2003uz, Pfeuffer:2008mz, Bakry:2014gea}: we first fix the spacetime trajectory of the point-like junction, then integrate over the NGB fluctuations around the saddle point determined by this trajectory, and finally perform the quantum-mechanical path-integral over the junction worldline. In the static gauge, the baryon junction worldline in $(d+1)$-dimensional spacetime
$W_\mu = (W_0,\vec{W})$ is parameterized as
\begin{equation}
    W_0=\tau,~~\vec{W}=l_s \vec{w}(\tau)=l_s(w_1(\tau),w_2(\tau),\dots,w_d(\tau))~.
\end{equation}
The trajectory $\vec{w}(\tau)$ sets the boundary profiles for the NGBs $x_i$, $y_i$, and $z_i$, as illustrated by the geometry in figure \ref{pic_baryon config}. We take the following convention that at the junction $\sigma=0$:
\begin{equation}
\label{eq_trajectory boundary condition}
\begin{aligned}
        &x_2=w_2~,~~y_2=-\frac{\sqrt{3}}{2}w_1-\frac{1}{2}w_2~,~~z_2=\frac{\sqrt{3}}{2}w_1-\frac{1}{2}w_2~,\\
        &\text{and}~~x_j=y_j=z_j=w_j~~\text{for}~~3\leq j\leq d.
\end{aligned}
\end{equation}
The saddle point of the NGB fields subject to the boundary condition \eqref{eq_trajectory boundary condition} and action \eqref{eq_baryon strings order 0} is solved by the free theory equation of motion. For convenience, we introduce the Dirichlet-Dirichlet Green's function on a cylinder $G(\sigma,{\sigma}',\tau;q)$ as in equation \eqref{eq_app_DD green function def}, and define the modular parameters $q_x$, $q_y$, and $q_z$ as
\begin{equation}
    q_x\equiv e^{-\frac{2\pi^2R}{L_x}}~,~~q_y\equiv e^{-\frac{2\pi^2R}{L_y}}~,~~\text{and}~~q_z\equiv e^{-\frac{2\pi^2R}{L_z}}~.
\end{equation}
The saddle point configurations of the NGBs $x_i^\star$, $y_i^\star$, and $z_i^\star$ then take the form
\begin{equation}
\label{eq_saddle point profile}
\begin{aligned}
    \big(x_2^\star,x_j^\star\big)={}&\int_{{\sigma}'=0} d {\tau}' \partial_{{\sigma}'}G(\sigma,{\sigma}',\tau-{\tau}';q_x)\big( w_2({\tau}'),w_j({\tau}')\big)~,\\
    \big(y_2^\star,y_j^\star\big)={}&\int_{{\sigma}'=0} d {\tau}' \partial_{{\sigma}'}G(\sigma,{\sigma}',\tau-{\tau}';q_y)\big( -\frac{\sqrt{3}}{2}w_1({\tau}')-\frac{1}{2}w_2({\tau}'),w_j({\tau}')\big)~,\\
    \big(z_2^\star,z_j^\star\big)={}&\int_{{\sigma}'=0} d {\tau}' \partial_{{\sigma}'}G(\sigma,{\sigma}',\tau-{\tau}';q_z)\big(\frac{\sqrt{3}}{2}w_1({\tau}')-\frac{1}{2}w_2({\tau}'),w_j({\tau}')\big)~.
\end{aligned}
\end{equation}

For free theories, the fluctuation spectrum is independent of the Gaussian saddle point around which one expands \cite{Cuomo:2021kfm, Cuomo:2024psk}. The path-integral therefore factorizes into two pieces: the partition function of three confining strings with Dirichlet boundary conditions on both ends, and the path-integral over the baryon junction worldline, governed by the saddle point action. Specifically, we find that
\begin{equation}
\label{eq_order 0 factorize}
    \mathcal{Z}^{(0)}=\frac{1}{\big(\eta(q_x)\eta(q_y)\eta(q_z)\big)^{d-1}}\int \mathcal{D}\vec{w}\exp{\left(-S^{(0)}_{\text{strings}}(x_i^\star,y_i^\star, z_i^\star)\right)}~.
\end{equation}
In equation \eqref{eq_order 0 factorize}, $S^{(0)}_{\text{strings}}(x_i^\star,y_i^\star, z_i^\star)$ denotes the order $0$ effective action \eqref{eq_baryon strings order 0} evaluated at the saddle point \eqref{eq_saddle point profile}. As a functional of the junction trajectory $\vec{w}(\tau)$, it gives the trapping potential arising from the backreaction of confining strings. See equations \eqref{eq_app_K def} and \eqref{eq_app_trapping potential} in Appendix \ref{sec_app_greens functions}, for the explicit form of $S^{(0)}_{\text{strings}}(x_i^\star,y_i^\star, z_i^\star)$. We carry out the Gaussian path-integral in equation \eqref{eq_order 0 factorize} and find that
\begin{equation}
\label{eq_partition function order 0 general}
\mathtoolsset{multlined-width=0.9\displaywidth}
\begin{multlined}
\mathcal{Z}^{(0)}=\frac{({L_x^{-1}+L_y^{-1}+L_z^{-1})^{1-\frac{d}{2}}}}{(2\pi R)^{\frac{d}{2}}\big(\eta(q_x)\eta(q_y)\eta(q_z)\big)^{d-1}}\sqrt{\frac{L_x L_y L_z}{L_x+L_y+L_z}}\prod_{n\in\mathbb{N}^+}\Bigg[\tanh{(\frac{nL_x}{R})}\\
  \times \tanh{(\frac{nL_y}{R})}\tanh{(\frac{nL_z}{R})}\frac{\left(\coth{(\frac{nL_x}{R})}+\coth{(\frac{nL_y}{R})}+\coth{(\frac{nL_z}{R})}\right)^{2-d}}{\tanh{(\frac{nL_x}{R})}+\tanh{(\frac{nL_y}{R})}+\tanh{(\frac{nL_z}{R})}}\Bigg]~.
\end{multlined}
\end{equation}

The same method extends to the order $1$ correction $\langle S^{(1)}_\text{junction}\rangle$, which is given by the quantum mechanical path-integral:
\begin{equation}
    \langle S^{(1)}_\text{junction}\rangle=\frac{Ml_s^2}{2}\frac{\int \mathcal{D}\vec{w}(\partial_\tau\vec{w})^2\exp{\left(-S^{(0)}_{\text{strings}}(x_i^\star,y_i^\star, z_i^\star)\right)}}{\int \mathcal{D}\vec{w}\exp{\left(-S^{(0)}_{\text{strings}}(x_i^\star,y_i^\star, z_i^\star)\right)}}~.
\end{equation}
We thereby find the following infinite sum representation for $\langle S^{(1)}_\text{junction}\rangle$
\begin{equation}
\label{eq_partition function order 1 general}
\mathtoolsset{multlined-width=0.9\displaywidth}
\begin{multlined}
    \langle S^{(1)}_\text{junction}\rangle=\frac{M l_s^2}{R}\sum_{n\in\mathbb{N}^+}\left[\frac{(d-2)n}{\coth{(\frac{nL_x}{R})}+\coth{(\frac{nL_y}{R})}+\coth{(\frac{nL_z}{R})}
    }+\frac{4n}{3}\tanh{(\frac{nL_x}{R})}\right.\hfill\\
 \hfill \left.\times \tanh{(\frac{nL_y}{R})}\tanh{(\frac{nL_z}{R})}\frac{\coth{(\frac{nL_x}{R})}+\coth{(\frac{nL_y}{R})}+\coth{(\frac{nL_z}{R})}}{\tanh{(\frac{nL_x}{R})}+\tanh{(\frac{nL_y}{R})}+\tanh{(\frac{nL_z}{R})}}\right]~.
\end{multlined}
\end{equation}

To the best of our knowledge, the partition function contributions \eqref{eq_partition function order 0 general} and \eqref{eq_partition function order 1 general} with generic $L_x$, $L_y$, and $L_z$ do not admit a closed form. See Appendix \ref{sec_app_rational points} for further discussion of cases in which the ratios of confining string lengths are rational. We note a particularly simple case: 
\begin{equation}
\label{eq_equilateral def}
    \text{equilateral baryon}~:~~L_x=L_y=L_z=L~,~~\text{with}~~q\equiv e^{-\frac{2\pi^2R}{L}}~.
\end{equation}
For equilateral baryons, the contributions \eqref{eq_partition function order 0 general} and \eqref{eq_partition function order 1 general} can be written in terms of (quasi-)modular forms of the parameters $q$ and $\sqrt{q}$ as follows:
\begin{equation}
\label{eq_equilateral baryon}
\begin{aligned}
          \text{equilateral baryon}~:~~&\mathcal{Z}^{(0)}=\frac{1}{(\eta(\sqrt{q}))^d(\eta(q))^{d-3}}~,~~\text{and}\\
       &\langle S_\text{junction}^{(1)}\rangle=\frac{(d+2)Ml_s^2}{144L}\ln{q} \left(2E_2(q)-E_2(\sqrt{q}) \right)~,
   \end{aligned} 
\end{equation}
in agreement with \cite{Komargodski:2024swh}.

\subsection{$S$-wave scattering}
\label{sec_s-wave scattering}

Having analyzed the open channel in the last section, we now turn to the closed channel representation \eqref{eq_baryon closed channel} of the baryon partition function. See also the discussion in \cite{Komargodski:2024swh}.

Higher-point functions of the Polyakov loop operators are governed by the potential $V(\Phi_a,\Phi_b,\dots)$ in the effective action \eqref{eq_closed string EFT}. For trivalent baryon junctions in figure \ref{pic_baryon config}, we have assumed that the underlying $(d+1)$-dimensional gauge theory is endowed with a $\mathbb{Z}_3$ $1$-form symmetry, either exact in the UV or emergent in the IR. Upon dimensional reduction, this $1$-form symmetry descends to a $\mathbb{Z}_3$ $0$-form acting on the complex scalar fields $\Phi_a$ in $d$-dimensional space. We note that the $\mathbb{Z}_3$-symmetric potential $V(\Phi_a,\Phi_b,\dots)$ admits cubic interaction vertices of the form:
\begin{equation}
\label{eq_cubic coupling def}
\mathtoolsset{multlined-width=0.9\displaywidth}
\begin{multlined}
    V(\Phi_a,\Phi_b,\dots)\supset \frac{1}{2}\left(\frac{3}{2\pi}\right)^{\frac{d}{2}} (\sqrt{\pi}l_s)^{\frac{d-3}{2}}\sum_{a,b,c} \sqrt{E^{\text{closed}}_aE^{\text{closed}}_bE^{\text{closed}}_c}\hfill\\
   \hfill  \times \Big[C_{abc}^{(0)}\Phi_a\Phi_b\Phi_c+C_{abc}^{(2)}\partial_{\nu} \Phi_a\partial_{\nu}\Phi_b\Phi_c+(\text{c.c.})+O\left(\partial^4\right)\Big]~,
\end{multlined}
\end{equation}
where $C^{(n)}_{abc}$ denote the order $n$ cubic coupling constants, fully symmetric in the massive mode indices $a$, $b$, $c$. The various prefactors in equation \eqref{eq_cubic coupling def} are chosen so that $C^{(0)}_{abc}$ are dimensionless and geometric factors are canceled, as we will see below. 

\begin{figure}[thb]
\centering
\includegraphics[width=\textwidth]{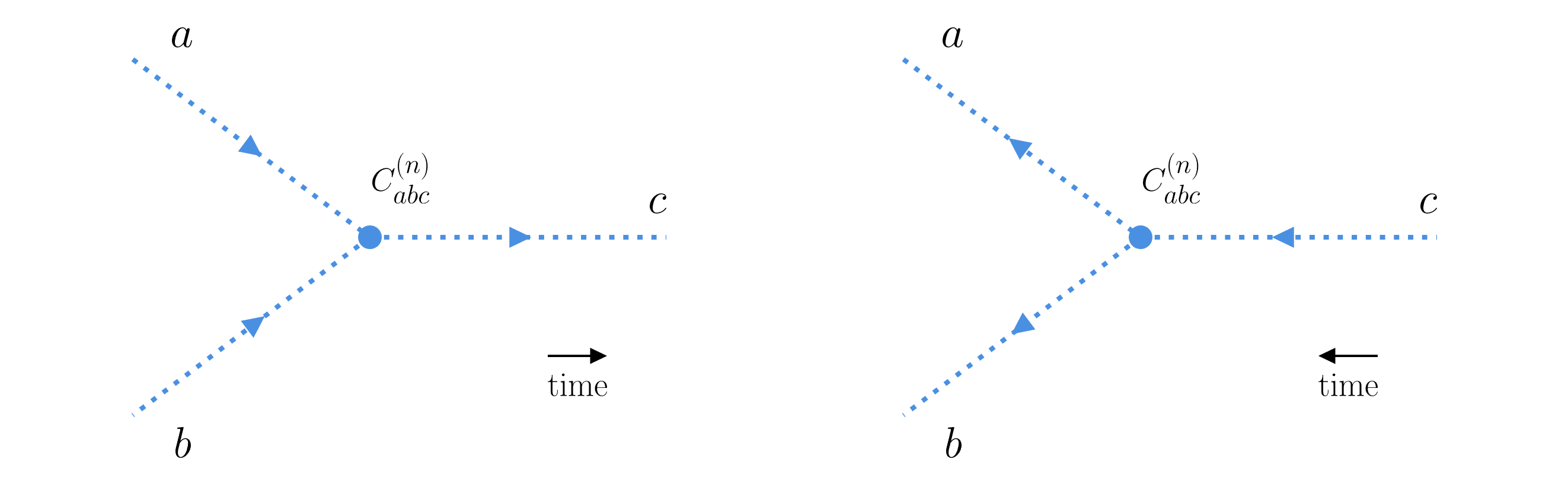}
  \caption{Tree-level scattering amplitudes of massive modes in the closed channel. Interpreting one direction of the $d$-dimensional space as time, the cubic interaction vertices in equation \eqref{eq_cubic coupling def} describe the $2\to 1$ fusion process (Left)
and the $1\to 2$ decay process (Right). The external legs in this figure are generally off-shell, since all modes in the closed channel have approximately the same mass \eqref{eq_closed energy level}. } \label{pic_scattering amp}
\end{figure}

At tree-level in the perturbation theory, these interaction vertices determine scattering amplitudes between massive particles in $d$-dimensional space. See also figure \ref{pic_scattering amp}. In particular, the coupling constants $C_{abc}^{(0)}$ are associated with the $s$-wave scattering. Likewise, $C_{abc}^{(2)}$ correspond to the $p$-wave scattering, while higher-spin scatterings (e.g.\ $d$-wave) arise at order $O(\partial^4)$ in the potential \eqref{eq_cubic coupling def}. We find that the following interaction vertices become total derivatives upon using the equations of motion:
\begin{equation}
2\sum_{a,b,c}C_{abc}^{(2)}\partial_{\nu} \Phi_a\partial_{\nu}\Phi_b\Phi_c=\sum_{a,b,c}C_{abc}^{(2)}\Big(\partial_{\nu}(\partial_{\nu} \Phi_a\Phi_b\Phi_c)-(\partial_{\nu}^2\Phi_a)\Phi_b\Phi_c\Big)~.
\end{equation}
As far as local kinetic terms are concerned, the coupling constants $C_{abc}^{(2)}$ can be removed by redefining $C^{(0)}_{abc}$ and the higher-spin terms. We therefore find that the tree-level amplitudes in the three-point function $\langle\Phi_a\Phi_b\Phi_c\rangle$ are determined entirely by $s$-wave scatterings up to order $3$.

As in Section \ref{sec_The open-closed duality}, we argue that the loop contributions in the three-point function \eqref{eq_baryon closed channel} are exponentially suppressed. The partition function up to the order $3$ is thereby given by the Feynman integral 
\begin{equation}
\label{eq_baryon closed channel Feynman integral}
\mathtoolsset{multlined-width=0.9\displaywidth}
\begin{multlined}
\mathcal{Z}=\sum_{a,b,c}\int d^d\vec{W}l_s^{-d}C_{abc}^{(0)}v_av_bv_c\Bigg(\frac{(E^{\text{closed}}_a)^{\frac{d}{2}}l_s^{d-1}}{\sqrt{\pi}(2L_{wx})^{\frac{d-2}{2}}}K_{\frac{d-2}{2}}\left(E^{\text{closed}}_a L_{wx}\right)\Bigg)\\
\times \Bigg(\frac{(E^{\text{closed}}_b)^{\frac{d}{2}}l_s^{d-1}}{\sqrt{\pi}(2L_{wy})^{\frac{d-2}{2}}}K_{\frac{d-2}{2}}\left(E^{\text{closed}}_b L_{wy}\right)\Bigg) \Bigg(\frac{(E^{\text{closed}}_c)^{\frac{d}{2}}l_s^{d-1}}{\sqrt{\pi}(2L_{wz})^{\frac{d-2}{2}}}K_{\frac{d-2}{2}}\left(E^{\text{closed}}_c L_{wz}\right)\Bigg)~,
\end{multlined}
\end{equation}
where $L_{wx}=|\vec{X}-\vec{W}|$, $L_{wy}=|\vec{Y}-\vec{W}|$, and $L_{wz}=|\vec{Z}-\vec{W}|$. In equation \eqref{eq_baryon closed channel Feynman integral}, we have employed the decomposition \eqref{eq_Polyakov decomposition} of the Polyakov operators and used the massive propagator \eqref{eq_2pt closed channel} derived from the effective action \eqref{eq_closed string EFT}. The Feynman integral is heavily dominated by its saddle point contribution, which we examine in detail in Section \ref{sec_Higher-loop corrections}. For later convenience, we introduce the dual modular parameters: 
\begin{equation}
    \tilde{q}_x\equiv e^{-\frac{2L_x}{R}}~,~~\tilde{q}_y\equiv e^{-\frac{2L_y}{R}}~,~~\text{and}~~\tilde{q}_z\equiv e^{-\frac{2L_z}{R}}~.
\end{equation}
The partition function then takes the form \cite{Komargodski:2024swh}
\begin{equation}
\mathtoolsset{multlined-width=0.9\displaywidth}
\begin{multlined}
\label{eq_closed_channel_final}
\mathcal{Z}=\left(\frac{3\pi R}{L_x+L_y+L_z}\right)^{d-\frac{3}{2}}
\left(\frac{(L_x+L_y+L_z)^2}{3(L_xL_y+L_yL_z+L_zL_x)}\right)^{\frac{d}{2}-1}
\\
\times\frac{e^{-2\pi R(L_x+L_y+L_z)/l_s^2}}{2^{\frac{d}{2}}(\tilde{q}_x\tilde{q}_y\tilde{q}_z)^{\frac{d-1}{24}}}
\sum_{a,b,c}\left(C^{(0)}_{abc}+O(\partial^2)\right)v_{a}v_{b}v_{c}
\tilde{q}_x^{n_{a}}\tilde{q}_y^{n_{b}}\tilde{q}_z^{n_{c}}~,
\end{multlined}
\end{equation}
where $n_a$, $n_b$, $n_c$ are the integer excitation levels given in equation \eqref{eq_closed energy level}.

The open-closed duality of the baryon junctions identifies the two representations \eqref{eq_perturbative expansion} and \eqref{eq_closed_channel_final} of the partition function. This allows us to fix the coupling constants $C^{(0)}_{abc}$ by consistency conditions, even though the underlying confining gauge theory is strongly coupled. Indeed, the open channel terms in equations \eqref{eq_partition function order 0 general} and \eqref{eq_partition function order 1 general} admit expansions in $\tilde{q}_x$, $\tilde{q}_y$, and $\tilde{q}_z$ using the modular transformation. Putting together $\mathcal{Z}^{(0)}$, $\langle S^{(1)}_{\text{junction}}\rangle$, and the classical contribution \eqref{eq_baryon classical contribution} to the partition function, we find the four coupling constants involving the closed string states $\textbf{0}$ and $\textbf{1}$: 
\begin{equation}
\label{eq_coupling const 1}
\mathtoolsset{multlined-width=0.9\displaywidth}
\begin{multlined}
C^{(0)}_{\textbf{0}\textbf{0}\textbf{0}}=e^{-2 \pi R M}\left[1+\frac{(d+2) M l_s^2}{36 R}+O\left(R^{-2}\right)\right]~,\hfill\\
C^{(0)}_{\textbf{0}\textbf{0}\textbf{1}}=\frac{e^{-2 \pi R M}}{3\sqrt{d-1}}\left[(d-3)+\frac{(d+2)(d+21)M l_s^2}{36 R}
+O\left(R^{-2}\right)\right]~,\hfill\\
C^{(0)}_{\textbf{0}\textbf{1}\textbf{1}}=\frac{e^{-2 \pi R M}}{9(d-1)}\left[(d^2-2 d+5)
+\frac{(d^3+48 d^2-143 d-278) M l_s^2}{36 R}
+O\left(R^{-2}\right)\right]~,\hfill\\
C^{(0)}_{\textbf{1}\textbf{1}\textbf{1}}=\frac{e^{-2 \pi R M }}{27(d-1)^{\frac{3}{2}}}\biggl[(d-3)(d^2+6d-19)\hfill\\
\hfill +\frac{(d^4+77d^3-319d^2+1495d-1470) M l_s^2}{36 R}
+O\left(R^{-2}\right)\biggr]~.
\end{multlined}
\end{equation}
We note that the coupling constants in \eqref{eq_coupling const 1} depend on the intrinsic EFT
parameters $l_s$ and $M$, as well as on the radius $R$ of Polyakov loops. They are independent of the operator positions (i.e., $L_x$, $L_y$, and $L_z$), as expected from the locality of the interaction vertices in \eqref{eq_cubic coupling def}. We will return to this point in Section \ref{sec_Higher-loop corrections}, where higher-order corrections are included.

\subsection{An accidental $\mathbb{Z}_2$ symmetry}
\label{sec_symmetry constraints}

We have shown that the interaction vertices in the effective action of propagating closed strings can be determined using open-closed duality. In particular, $s$-wave scatterings of these closed strings are associated with the scale $\Lambda_{\text{$s$-wave}}$ that generally obeys
\begin{equation}
    (\Lambda_{\text{$s$-wave}})^{3-\frac{d}{2}}~\sim~ l_s^{\frac{d-3}{2}}\sqrt{E^{\text{closed}}_aE^{\text{closed}}_bE^{\text{closed}}_c}C_{abc}^{(0)}~\sim ~R^{\frac{3}{2}}l_s^{\frac{d-9}{2}}e^{-2\pi R M}~.
\end{equation}
Notably, the baryon junction mass $M$ sets a non-perturbative factor $e^{-2\pi R M}$ that governs the interactions between long strings. The perturbative description applies to the effective action \eqref{eq_closed string EFT} given that $M\geq 0$.  Nevertheless, consistency by itself places no strict bounds on the junction mass $M$. See \cite{Komargodski:2024swh} for a relevant discussion.
\begin{figure}[thb]
\centering
\includegraphics[width=.9\textwidth]{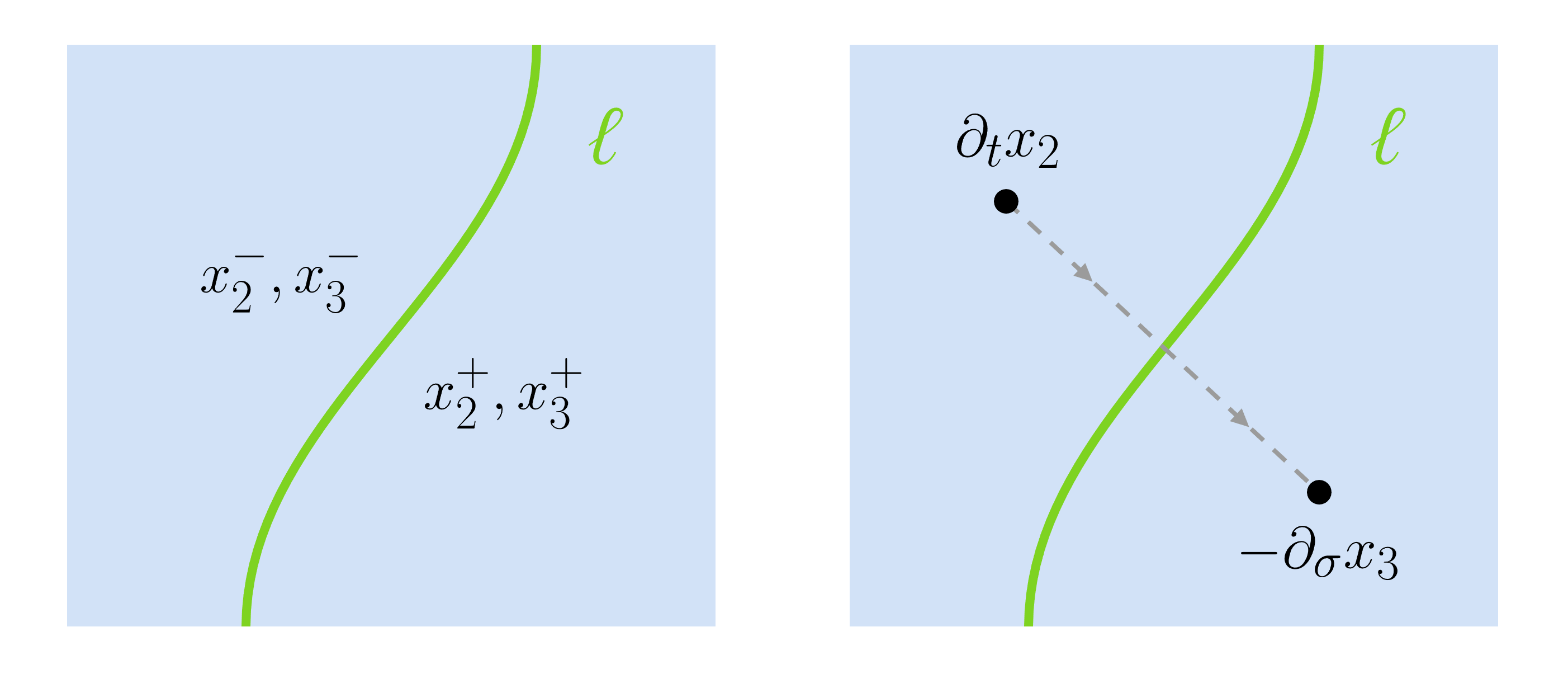}
  \caption{Topological defect line $\mathcal{D}$ on the confining string worldsheet. Left: the NGB field profiles are taken to be discontinuous across the line $\ell$, and the defect action is given by the bilinear coupling between $x_{i}^-$ and $x_{i}^+$. Right: local operators transform according to equation \eqref{eq_symmetry rule} upon crossing the defect line. The $\mathbb{Z}_2$ symmetry transformation rule can be schematically written as $(\partial_tx_2,\partial_\sigma x_2)\xrightarrow{\mathcal{D}} (-\partial_\sigma x_3,-\partial_t x_3)\xrightarrow{\mathcal{D}} (\partial_tx_2,\partial_\sigma x_2)$.} \label{pic_symmetry defect}
\end{figure}

Interestingly, the effective string theory in $(3+1)$-dimensional spacetime (i.e., $d=3$) presents an accidental symmetry that is broken by the junction mass $M$. To see this, we consider a class of line defects living on the confining string worldsheet with the effective action \eqref{eq_NG action 3}. The defect $\mathcal{D}$ is defined by coupling the NGB fields across the line manifold $\ell$ as follows:
\begin{equation}
\label{eq_duality defect}
    \mathcal{D}(\ell )\equiv \exp{\left(i\int_\ell (x_2^-dx_3^+-x_3^-dx_2^+)\right)}=\exp\left(i S^{(0)}_{\text{defect}}\right)~,
\end{equation}
where $x_{i}^-$ denotes the field profile on the left-hand side
of $\ell$, while $x_{i}^+$ denotes the profile on the right-hand side. See also figure \ref{pic_symmetry defect}. This construction is analogous to the T-duality defects in the 2-dimensional free compact boson theory \cite{Kapustin:2009av,Choi:2021kmx,Niro:2022ctq,Shao:2025qvf}. The insertion of the defect $\mathcal{D}$ preserves the transverse rotation group $SO(2)$ and modifies the equations of motion for the NGB fields. It follows from the variational principle that the fields on the two sides of the defect are matched according to
\begin{equation}
\label{eq_symmetry rule}
    \big(\partial_t x_2^-,\partial_\sigma x_2^-,\partial_t x_3^-,\partial_\sigma x_3^-\big)=\big(\partial_\sigma x_3^+,\partial_t x_3^+,-\partial_\sigma x_2^+,-\partial_t x_2^+\big)~.
\end{equation}
For the probe baryon configurations shown in figure \ref{pic_baryon config}, we can similarly define defects $\mathcal{D}$ on the confining string worldsheets using variables $y_i^\pm$ and $z_i^\pm$, respectively.

The free theory limit of the baryon effective action \eqref{eq_baryon action} is governed by the quadratic kinetic term \eqref{eq_baryon strings order 0} and the junction condition \eqref{eq_junction conditions 2}. One readily verifies that the worldsheet stress-energy tensor is continuous across $\mathcal{D}$ in this limit. Equation \eqref{eq_duality defect} therefore defines a topological defect line in the free theory, which generates a $\mathbb{Z}_2$ global symmetry on the worldsheet. In particular, the symmetry transformation rule for local operators is given by \eqref{eq_symmetry rule}.\footnote{Local vertex operators (e.g., $\exp{(ik_ix_i)}$) are mapped to semi-local twist operators under this $\mathbb{Z}_2$ symmetry. These twist operators are characterized by the field monodromy $x_i\to x_i+\text{const}$ around the operator insertion.}

\begin{figure}[thb]
\centering
\includegraphics[width=\textwidth]{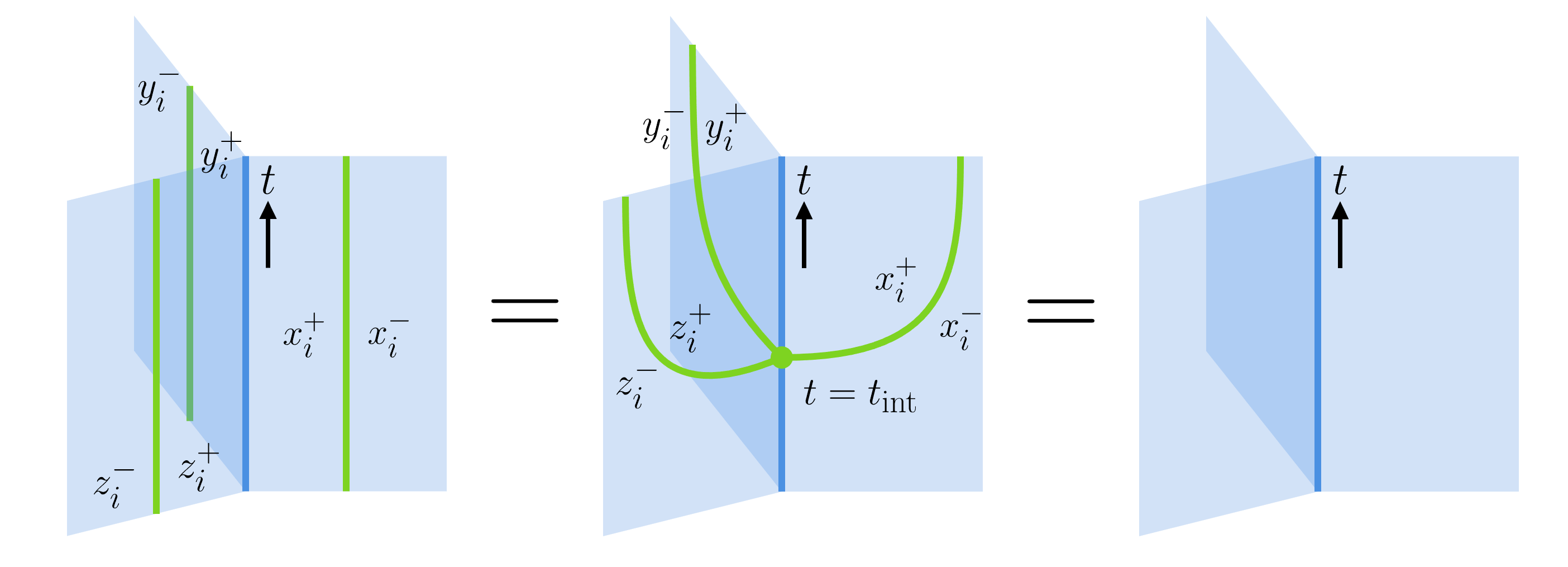}
  \caption{Merging the topological defect lines with the baryon junction worldline.} \label{pic_junction fusion}
\end{figure}

Notably, the baryon junction condition \eqref{eq_junction conditions 2} preserves the $\mathbb{Z}_2$ symmetry generated by the topological defect line $\mathcal{D}$.\footnote{For effective string theories in $(d+1)$-dimensional spacetime, we can analogously define a line defect that preserves the transverse rotation group $SO(d-1)$ as follows:
\begin{equation*}
    {\mathcal{D}}'(\ell )\equiv \exp{\left(i\int_\ell x_i^-dx_i^+\right)}~,~~\text{where}~~2\leq i \leq d~.
\end{equation*}
The baryon junction condition \eqref{eq_junction conditions 2}, however, is not invariant under the symmetry generated by ${\mathcal{D}}'$.} The key observation is as follows: the linear combinations of NGB fields $\xi_2^{[1]}$, $\xi_3^{[2]}$, and $\xi_3^{[3]}$ satisfy Dirichlet boundary conditions at the junction, while the other combinations $\xi_2^{[2]}$, $\xi_2^{[3]}$, and $\xi_3^{[1]}$ satisfy Neumann boundary conditions. The $\mathbb{Z}_2$ symmetry \eqref{eq_duality defect} exchanges Dirichlet and Neumann conditions and acts by a $\pi/2$ rotation on the spatial indices transverse to the confining string, therefore leaving the junction condition invariant. In the free theory limit, the baryon junction worldline can be merged with the defect $\mathcal{D}$ across the three connected worldsheets as in figure \ref{pic_junction fusion}. The $\mathbb{Z}_2$ symmetry acts on operators localized at the baryon junction through the topological intersection of the junction worldline with the line defects. 

We now examine whether the $\mathbb{Z}_2$ symmetry generated by the defect $\mathcal{D}$ persists in the full effective theory \eqref{eq_baryon action}. On the confining string worldsheets, we note that the quartic interaction in equation \eqref{eq_baryon strings order 2} is invariant under the $\mathbb{Z}_2$ transformation \eqref{eq_symmetry rule}. The leading symmetry breaking interaction on the worldsheets arises at order $4$ and is controlled by the Polyakov–Kleinert rigidity term.

On the other hand, the baryon junction worldline preserves the $\mathbb{Z}_2$ symmetry if its intersection with defect $\mathcal{D}$ remains topological. We first note that coupling the worldsheets on the two sides of the defect using \eqref{eq_duality defect} requires the longitudinal displacements $\delta X_1$, $\delta Y_1$, and $\delta Z_1$ in \eqref{eq_longitudinal displacement} to be continuous across the defect line. This leads to the following consistency condition at the intersection point $t=t_{\text{int}}$ (see figure \ref{pic_junction fusion}):
\begin{equation}
\label{eq_gluing condition}
    \text{gluing condition}~:~x_2^+=x_2^-~,~y_2^+=y_2^-~,\text{ and }z_2^+=z_2^-\text{ at }(t,\sigma)=(t_\text{int},0)~,
\end{equation}
which can be implemented using Lagrange multipliers at the intersection point.

We consider the worldline stress-energy tensor $\tilde{T}$ \cite{Cuomo:2021rkm}, defined by the local divergence of the worldsheet stress-energy tensors $T_{\upalpha\upbeta}[x_i]$, $T_{\upalpha\upbeta}[y_i]$, and $T_{\upalpha\upbeta}[z_i]$ as follows:
\begin{equation}
    \partial^\upalpha T_{\upalpha t}[x_i]+\partial^\upalpha T_{\upalpha t}[y_i]+\partial^\upalpha T_{\upalpha t}[z_i]=-\delta(\sigma)\partial_t\tilde{T}~.
\end{equation}
The operator $\tilde{T}$ represents the energy stored at the baryon junction, which vanishes in the free theory limit. In the full effective theory, $\tilde{T}$ receives corrections from the actions \eqref{eq_baryon junction order 1}, \eqref{eq_baryon displacement order 1}, \eqref{eq_baryon displacement order 2}, and other higher-order terms. For example, the contribution from longitudinal displacements on the $X$-string follows from \eqref{eq_displacement def} and takes the form
\begin{equation}
\label{eq_local tensor part}
    \tilde{T}\supset \sum_{n\in \mathbb{N}}(-\delta X_1)^{n+1}\partial_\sigma^{n}\left(\frac{1}{l_s^2}+\frac{1}{2}\left((\partial_t x_i)^2+(\partial_\sigma x_i)^2\right)+O(\partial^4)\right)~.
\end{equation}
Using the transformation rule \eqref{eq_symmetry rule} and the condition \eqref{eq_gluing condition}, we find that the corrections to the stress-energy tensor $\tilde{T}$ in \eqref{eq_local tensor part} are continuous across the intersection point $t=t_\text{int}$ up to order $2$. The discontinuity at $t=t_{\text{int}}$ arises from the worldline kinetic term \eqref{eq_baryon junction order 1} and is given by
\begin{equation}
\label{eq_discontinuity}
\mathtoolsset{multlined-width=0.9\displaywidth}
\begin{multlined}
\tilde{T}^+-\tilde{T}^- =\frac{Ml_s^2}{18}\Big[(\partial_tx_3^++\partial_ty_3^++\partial_tz_3^+)^2-(\partial_\sigma x_2^++\partial_\sigma y_2^++\partial_\sigma z_2^+)^2\hfill\\
\phantom{\tilde{T}\big|_{t>t_\text{int}}-\tilde{T}\big|_{t<t_\text{int}}=} +(\partial_tx^+_2+\partial_ty^+_2-2\partial_tz^+_2)^2-(\partial_\sigma x^+_3+\partial_\sigma y^+_3-2\partial_\sigma z^+_3)^2\hfill
\\
\hfill+3(\partial_tx^+_2-\partial_ty^+_2)^2-3(\partial_\sigma x^+_3-\partial_\sigma y^+_3)^2\Big]+O\left(\partial^4\right)~,
\end{multlined}
\end{equation}
where $\tilde{T}^+$ and $\tilde{T}^-$ denote the stress-energy tensor evaluated on the two sides of the intersection point. The discontinuity \eqref{eq_discontinuity} quantifies the obstruction to moving the intersection point topologically along the baryon junction worldline, thereby signaling the $\mathbb{Z}_2$ symmetry breaking.

We conclude that the baryon junction mass $M$ is the leading parameter that breaks the $\mathbb{Z}_2$ symmetry \eqref{eq_duality defect} in the $(3+1)$-dimensional effective string theory. Unlike the boundary parameters in \eqref{eq_Dirichlet action} and \eqref{eq_Neumann action}, the baryon junction mass can be defined without reference to external probes (e.g., static quarks) and is therefore intrinsic to the underlying UV theory. If the $\mathbb{Z}_2$ symmetry is exactly realized in the UV theory, it imposes strong constraints on the baryon junction mass, forcing $M=0$.

In the closed channel, the $\mathbb{Z}_2$ symmetry \eqref{eq_duality defect} constrains the interaction vertices in the effective potential $V(\Phi_a, \Phi_b,\dots )$ and implies selection rules for scattering processes. For example, the closed string ground state $\textbf{0}$ is even under the $\mathbb{Z}_2$ transformation, while the excited state $\textbf{1}$ is odd. By the $\mathbb{Z}_2$ symmetry, the coupling constants in \eqref{eq_cubic coupling def} must satisfy
\begin{equation}
\label{eq_selection rule}C^{(0)}_{\textbf{0}\textbf{0}\textbf{1}}=C^{(0)}_{\textbf{1}\textbf{1}\textbf{1}}= 0~,
\end{equation}
in agreement with the explicit results \eqref{eq_coupling const 1} with $d=3$ and $M=0$. We will examine these symmetry constraints further in Section \ref{sec_Higher-loop corrections}, where nonlinear corrections are taken into account. Finally, we note that the $s$-wave scatterings of closed strings in $(3+1)$-dimensional spacetime are associated with different scales. For a generic baryon junction $M$, the $\mathbb{Z}_2$ symmetry preserving (SP) and symmetry breaking (SB) scales are respectively
\begin{equation}
    \Lambda^{\text{SP}}_{\text{$s$-wave}}\sim Rl_s^{-2}e^{-\frac{4\pi }{3}RM}~,~~\text{and}~~\Lambda^{\text{SB}}_{\text{$s$-wave}}\sim M^{\frac{2}{3}}R^{\frac{1}{3}}l_s^{-\frac{2}{3}}e^{-\frac{4\pi }{3}RM}~.
\end{equation}

\section{Nonlinear corrections}
\label{sec_Higher-loop corrections}

In this section, we extend the open-closed duality analysis of the baryon junction to the next order in the effective string theory. To keep the discussion streamlined, we have moved the technical details of the calculations in this section to Appendix \ref{sec_app_greens functions}. 

In the open channel \eqref{eq_baryon open channel}, the order $2$ contributions to the probe baryon partition function include $\langle S^{(2)}_{\rm displacement}\rangle$, $\langle S^{(2)}_{\rm strings}\rangle$, $\langle (S^{(1)}_{\rm junction})^2\rangle$, and $\langle (S^{(1)}_{\rm displacement})^2\rangle$. These terms represent nonlinear corrections to the NGB fields $x_i$, $y_i$, and $z_i$, which can be computed using worldsheet loop diagrams (see figure \ref{pic_feynman diagram}). We present these results in Section \ref{sec_regularized partition function}.

In the closed channel \eqref{eq_baryon closed channel}, the partition function remains dominated by $s$-wave scattering and is given by the tree-level Feynman integral \eqref{eq_baryon closed channel Feynman integral} up to order $2$. By evaluating the integral \eqref{eq_baryon closed channel Feynman integral} perturbatively around its saddle point, we identify the $O(R^{-2})$ corrections to the coupling constants \eqref{eq_coupling const 1}. We elaborate on this computation in Section \ref{sec_Saddle-point approximation}.

Throughout this section, we will focus on the equilateral baryon configuration \eqref{eq_equilateral def}. This allows us to express the results in terms of (quasi-)modular forms of $q=\exp{(-2\pi^2 R/L)}$ and $\tilde{q}=\exp{(-2L/R)}$. Nevertheless, our analysis generalizes straightforwardly to other probe baryon configurations.

\subsection{Regularized corrections to the partition function }
\label{sec_regularized partition function}

\begin{figure}[thb]
\centering
\includegraphics[width=\textwidth]{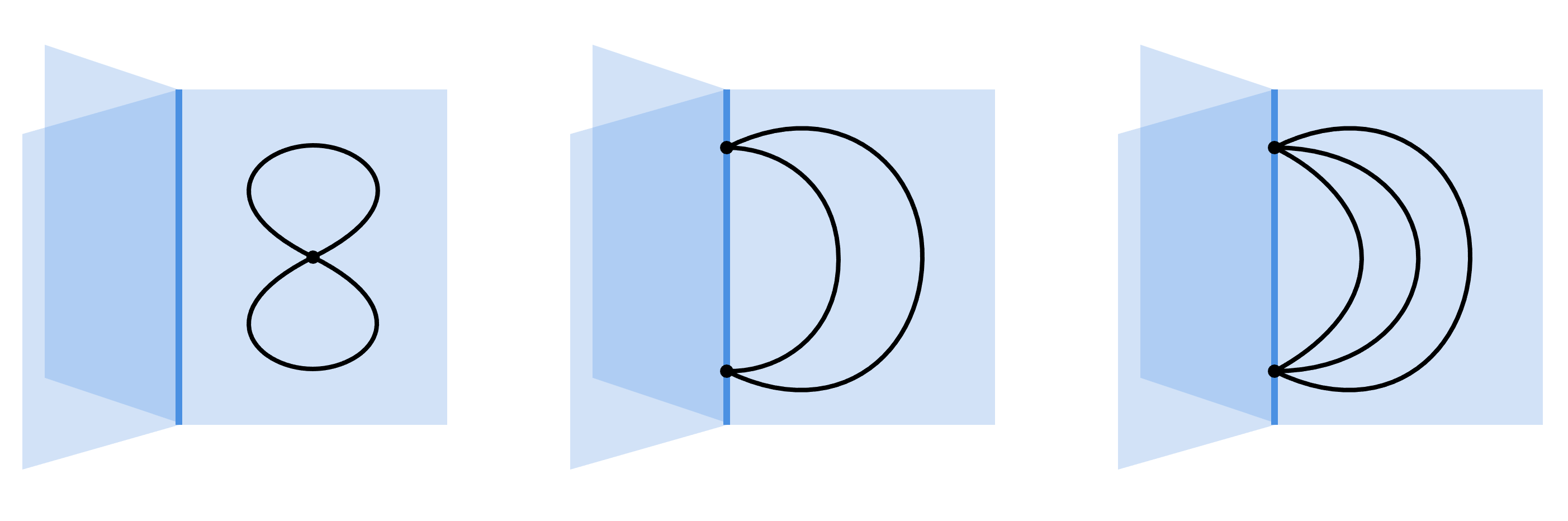}
  \caption{Worldsheet loop diagrams. The black lines denote the propagators of the NGB fields $x_i$, $y_i$, and $z_i$. See equations \eqref{eq_app_DD green function def}, \eqref{eq_app_ND green function def}, \eqref{eq_app_propagators 1}, and \eqref{eq_app_propagators 2} in Appendix \ref{sec_app_greens functions} for the explicit forms of these propagators. Left: $\langle S^{(2)}_{\rm strings}\rangle$ is given by the 2-loop diagrams on the confining string worldsheets; Middle: $\langle (S^{(1)}_{\rm junction})^2\rangle$ is given by the 1-loop diagrams on the baryon junction worldline; Right: $\langle (S^{(1)}_{\rm displacement})^2\rangle$ is given by the 2-loop diagrams on the baryon junction worldline.} \label{pic_feynman diagram}
\end{figure}

As shown in figure \ref{pic_feynman diagram}, the order 2 nonlinear corrections to the partition function \eqref{eq_perturbative expansion} are evaluated by loop diagrams either on confining string worldsheets or localized on the baryon junction worldline. These diagrams correspond to the coincident-point limits of various Green's functions and their integrals, which we review in Appendix \ref{sec_app_greens functions}. In general, the loop integrals in $\langle S^{(2)}_{\rm displacement}\rangle$, $\langle S^{(2)}_{\rm strings}\rangle$, $\langle (S^{(1)}_{\rm junction})^2\rangle$, and $\langle (S^{(1)}_{\rm displacement})^2\rangle$ are UV divergent. We adopt the zeta-function regularization and extract the finite contribution from these loop integrals.

We start with the quartic interaction \eqref{eq_baryon displacement order 2} at the baryon junction arising from longitudinal displacements. For the equilateral baryon configuration \eqref{eq_equilateral def}, we find by explicit calculation that
\begin{equation}
\label{eq_nonlinear 1}
    \langle S_\text{displacement}^{(2)}\rangle=0~.
\end{equation}
See also equations \eqref{eq_app_nonlinear 1} and \eqref{eq_app_triple derivatives}. With zeta-function regularization, equation \eqref{eq_nonlinear 1} also holds for general configurations of the probe baryon as in figure \ref{pic_baryon config}.

The quartic interaction \eqref{eq_baryon strings order 2} on the confining string worldsheets follows from the Nambu--Goto action. Using zeta-function regularization, we obtain $\langle S^{(2)}_{\rm strings}\rangle$ in terms of the Eisenstein series \eqref{eq_app_Eisenstein} as follows
\begin{equation}
\label{eq_nonlinear 2}
\mathtoolsset{multlined-width=0.9\displaywidth}
\begin{multlined}
\langle S_\text{strings}^{(2)}\rangle=\frac{\pi l_s^2}{13824 L^2}(\ln q)\Big[2(d+2) E_4(\sqrt{q})+4(d-3)(d-9)\big((E_2(q)\big)^2\hfill\\
\hfill+24(d-3)E_4(q) +(d^2-2 d-4)(E_2(\sqrt{q}))^2+4d(d-3) E_2(\sqrt{q}) E_2(q)\Big]\\
\phantom{\langle S_\text{strings}^{(2)}\rangle}=\frac{(4d^2-20d+23)l_s^2}{24\pi R^2(\ln\tilde{q})}+\frac{l_s^2}{144\pi R^2}\Big[(d-3)(2d-9)E_2(\tilde{q})\hfill\\
\phantom{\langle S_\text{strings}^{(2)}\rangle=}+2(2d^2-5d-4)E_2(\tilde{q}^2)\Big]+\frac{l_s^2}{3456\pi R^2}(\ln\tilde{q})\Big[8(d+2)E_4(\tilde{q}^2)\hfill\\
\phantom{\langle S_\text{strings}^{(2)}\rangle=}+6(d-3)E_4(\tilde{q})+4(d^2-2d-4)\big(E_2(\tilde{q}^2)\big)^2\hfill\\
+(d-3)(d-9)\big(E_2(\tilde{q})\big)^2+4d(d-3)E_2(\tilde{q})E_2(\tilde{q}^2)\Big]~.
\end{multlined}
\end{equation}
See also equations \eqref{eq_app_nonlinear 2}, \eqref{eq_app_int 1}, \eqref{eq_app_int 2}, and \eqref{eq_app_int 3}. In equation \eqref{eq_nonlinear 2}, we present two equivalent expressions for the regularized nonlinear correction $\langle S_\text{strings}^{(2)}\rangle$, written in terms of $q=\exp{(-2\pi^2 R/L)}$ and $\tilde{q}=\exp{(-2L/R)}$, respectively. These two expressions are mapped into each other by the modular transformation \eqref{eq_app_modular trans Eisenstein}, and they agree with the open-closed duality: In the open channel, equation \eqref{eq_nonlinear 2} represents subleading corrections to the probe baryon energy levels $E_{\text{baryon}}$. After weighting by the Boltzmann factor $\exp(-\beta E_{\text{baryon}})$, these corrections scale with positive powers of $\ln{q}\sim \beta$. On the other hand, equation \eqref{eq_nonlinear 2}  splits into terms that scale as $(\ln \tilde{q})^{-1}$, $(\ln \tilde{q})^{0}$ and $(\ln \tilde{q})^{1}$ in the dual variable. As we will show shortly,  the $(\ln \tilde{q})^{0}$ term in \eqref{eq_nonlinear 2} encodes corrections to the local interaction vertices $C_{abc}^{(0)}$ in the closed channel, whereas the $(\ln \tilde{q})^{-1}$ and $(\ln \tilde{q})^{1}$ terms follow from the tree-level Feynman integral \eqref{eq_baryon closed channel Feynman integral}.

The kinetic action of the baryon junction \eqref{eq_baryon junction order 1} generates the following nonlinear corrections at order $2$:
\begin{equation}
\label{eq_nonlinear 3}
\mathtoolsset{multlined-width=0.9\displaywidth}
\begin{multlined}
\langle(S_\text{junction}^{(1)})^2\rangle=\frac{(d+6) M^2l_s^4}{216 L^2} (\ln q)\Big[2 E_2(q)-E_2(\sqrt{q})\Big]+\frac{M^2l_s^4}{20736L^2}(\ln q)^2\hfill\\
\phantom{\langle(S_\text{junction}^{(1)})^2\rangle=}\times \Big[2 (d+6) E_4(\sqrt{q})-8(d+6)E_4(q)+(d-2) (d+4) (E_2(\sqrt{q}))^2\\
\hfill -4 (d+2)^2 E_2(\sqrt{q})E_2(q)+4 (d^2+6d+16) (E_2(q))^2\Big]\\
\phantom{\langle(S_\text{junction}^{(1)})^2\rangle}=\frac{M^2l_s^4}{1296R^2}\Big[8(d+6)E_4(\tilde{q}^2)-2(d+6)E_4(\tilde{q})-4(d+2)^2 E_2(\tilde{q}) E_2(\tilde{q}^2)\hfill\\
 \hfill   +(d^2+6 d+16)\bigl(E_2(\tilde{q})\bigr)^2+4(d-2)(d+4)\bigl(E_2(\tilde{q}^2)\bigr)^2\Big]~.
\end{multlined}
\end{equation}
See also equations \eqref{eq_app_nonlinear 3}, \eqref{eq_app_int 4}, and \eqref{eq_app_int 5}. Equation \eqref{eq_nonlinear 3} scales as $(\ln {\tilde{q}})^0$ in the variable $\tilde{q}$, indicating that the baryon junction mass $M$ affects only the local interaction vertices in the effective action \eqref{eq_closed string EFT}. We will return to this point in Section \ref{sec_Saddle-point approximation}.

Finally, the longitudinal displacements \eqref{eq_baryon displacement order 1} gives rise to the nonlinear correction 
\begin{equation}
\label{eq_nonlinear 4}
\mathtoolsset{multlined-width=0.9\displaywidth}
\begin{multlined}
\langle(S_\text{displacement}^{(1)})^2\rangle=\frac{\pi l_s^2}{1728L^2}(\ln q)\hfill\\
\hfill\times \Big[(E_2(\sqrt{q}))^2+4(d-3)(E_2(q))^2-E_4(\sqrt{q})-4(d-3)E_4(q)\Big]\\
\phantom{\langle(S_\text{displacement}^{(1)})^2\rangle}=\frac{(d-2)l_s^2}{3\pi R^2(\ln\tilde{q})}+\frac{l_s^2}{18\pi R^2}\Big[(d-3)E_2(\tilde{q})+2E_2(\tilde{q}^2)\Big]+\frac{l_s^2}{432\pi R^2}(\ln\tilde{q})\hfill\\
\phantom{\langle(S_\text{displacement}^{(1)})^2\rangle=}\times \Big[4\big(E_2(\tilde{q}^2)\big)^2+(d-3)\big(E_2(\tilde{q})\big)^2-(d-3)E_4(\tilde{q})-4E_4(\tilde{q}^2)\Big]~.
\end{multlined}
\end{equation}
See equation \eqref{eq_app_nonlinear 4}, \eqref{eq_app_int 6}, \eqref{eq_app_int 7} and \eqref{eq_app_int 8}. Notably, equation \eqref{eq_nonlinear 4} contains contributions scaling as $(\ln \tilde{q})^{-1}$, $(\ln \tilde{q})^{0}$ and $(\ln \tilde{q})^{1}$ when written in terms of the variable $\tilde{q}$, which differs from the corrections on the baryon junction worldline $\langle S_\text{junction}^{(1)}\rangle$ and $\langle(S_\text{junction}^{(1)})^2\rangle$.

\subsection{Saddle-point approximation}
\label{sec_Saddle-point approximation}

In Section \ref{sec_s-wave scattering}, we argued that the baryon partition function \eqref{eq_perturbative expansion} can be evaluated by the tree-level Feynman integral \eqref{eq_baryon closed channel Feynman integral}, which describes the $s$-wave scattering in the closed channel. We now compute the integral via the saddle-point approximation and include the nonlinear corrections at order $2$.

The saddle point is given by the $d$-dimensional vector $\vec{W}=(W_1,W_2,\dots)$ that maximizes the integrand of equation \eqref{eq_baryon closed channel Feynman integral}. Finding such a vector reduces to the Fermat--Weber problem, which we solve perturbatively in powers of $R^{-1}\sim \beta^{-1}$. Denoting the energy levels of the closed string states in \eqref{eq_closed energy level} by $n_a$, $n_b$, and $n_c$, we find that 
\begin{equation}
\label{eq_Fermat point}
    \vec{W}_\text{saddle}=\frac{Ll_s^2}{3 \pi R^2}\bigl(2 n_{a}-n_{b}-n_{c},\sqrt{3}(n_{b}-n_{c}),0,\dots \bigr)+O\left(R^{-3}\right)~.
\end{equation}
Around the saddle point \eqref{eq_Fermat point}, the $s$-wave scattering amplitude \eqref{eq_baryon closed channel Feynman integral} reduces to a Gaussian integral. It is convenient to introduce $\vec{W}=\vec{W}_\text{saddle}+\delta\vec{W}$, such that the partition function takes the form
\begin{equation}
\label{eq_closed channel partition function 1}
\mathtoolsset{multlined-width=0.9\displaywidth}
\begin{multlined}
\mathcal{Z}=\sum_{a,b,c}C_{abc}^{(0)}v_av_bv_cF_{abc}l_s^{-d}\int d^d\vec{W}\exp{(-\lambda_{ij}\delta W_i\delta W_j)}\\
    \times\left(1-\lambda_{ijkl}\delta W_i\delta W_j\delta W_k\delta W_l+\frac{1}{2}\left(\lambda_{ijk}\delta W_i\delta W_j\delta W_k\right)^2+O\left(\partial^4\right)\right)~,
\end{multlined}
\end{equation}
where $1\leq i,j,k,l\leq d$. In equation \eqref{eq_closed channel partition function 1}, $F_{abc}$ denotes the maximum of the integrand \eqref{eq_baryon closed channel Feynman integral} (i.e., the saddle point value), and it reads
\begin{equation}
\mathtoolsset{multlined-width=0.9\displaywidth}
\begin{multlined}
    F_{abc}=\left(\frac{3}{2\pi}\right)^{\frac{d}{2}}\left(\frac{\pi R}{L}\right)^{\frac{3}{2}(d-1)}\frac{e^{-6\pi RL/l_s^2}}{2\tilde{q}^{\frac{d-1}{8}}}\tilde{q}^{n_a+n_b+n_c}\bigg\{ 1-\frac{l_s^2}{2\pi R^2}\bigg[\frac{3(d-3)(d-1)}{4\ln\tilde{q}}\hfill\\
    +\frac{(d-1)}{8}\big(d-1-8(n_a+n_b+n_c)\big)+\Big(\frac{(d-1)^2}{64}-\frac{(d-1)}{4}(n_a+n_b+n_c)\\
    \hfill+6(n_a^2+n_b^2+n_c^2)-(n_a+n_b+n_c)^2\Big)\frac{\ln \tilde{q}}{3}\bigg]+O\left(\partial^4\right)\bigg\}~.\\
\end{multlined}
\end{equation}
The coefficients $\lambda_{ij}$, $\lambda_{ijk}$, and $\lambda_{ijkl}$ are fully symmetric in the spatial indices, and the non-zero coefficients are listed in table \ref{tab_saddle pt coeff}.
{\renewcommand{\arraystretch}{1.3}
\begin{table}[h!]
\centering
\begin{tabular}{c|c}
 & $\lambda \times (-\frac{\ln{\tilde{q}}}{3\pi})$\\
  \hline
    $\lambda_{11}$ & $\frac{1}{l_s^2}+\big(n_b+n_c-n_a-\frac{d-1}{24}\big)\frac{1}{\pi R^2}$\\
    $\lambda_{22}$ & $\frac{1}{l_s^2}+\big(\frac{5}{3}n_a-\frac{1}{3}n_b-\frac{1}{3}n_c-\frac{d-1}{24}\big)\frac{1}{\pi R^2}$\\
    $\lambda_{12}$ & $(n_b-n_c)\frac{2}{\sqrt{3}\pi R^2}$\\
    $\lambda_{jj}$ & $\frac{2}{l_s^2}+\big(\frac{2}{3}(n_a+n_b+n_c)-\frac{d-1}{12}-\frac{d-1}{\ln{\tilde{q}}}\big)\frac{1}{\pi R^2}$
\end{tabular}\quad\quad
\begin{tabular}{c|c}
 & $\lambda \times (-\frac{\ln{\tilde{q}}}{3\pi})$\\
  \hline
    $\lambda_{111}$ & $\frac{1}{(\ln \tilde{q})l_s^2R}$\\
    $\lambda_{122}$& $-\frac{1}{(\ln \tilde{q})l_s^2R}$\\
    $\lambda_{1111}$, $\lambda_{2222}$ & $\frac{1}{4(\ln\tilde{q})^2l_s^2R^2}$\\
    $\lambda_{1122}$ & $\frac{1}{12(\ln\tilde{q})^2l_s^2R^2}$\\
    $\lambda_{11jj}$, $\lambda_{22jj}$ & $\frac{1}{3(\ln\tilde{q})^2l_s^2R^2}$\\
    $\lambda_{jjjj}$ & $-\frac{2}{(\ln\tilde{q})^2l_s^2R^2}$\\
    $\lambda_{jjkk}$ ($j\neq k$) & $-\frac{2}{3(\ln\tilde{q})^2l_s^2R^2}$
\end{tabular}\quad \quad \quad \quad
\caption{Saddle point coefficients of the Feynman integral. In this table, we take $3\leq j,k\leq d$. The coefficients
are given in units of $(-\frac{\ln{\tilde{q}}}{3\pi})=\frac{2L}{3\pi R}$, and we retain terms up to the order $O(\partial^2)$.\label{tab_saddle pt coeff}}
\end{table}}

 Carrying out the Gaussian integral \eqref{eq_closed channel partition function 1}, we find that
\begin{equation}
\label{eq_closed channel partition function 2}
\mathtoolsset{multlined-width=0.9\displaywidth}
\begin{multlined}
    \mathcal{Z}=\left(\frac{\pi R}{L}\right)^{d-\frac{3}{2}} \frac{e^{-6 \pi R L / l_s^2}}{2^{\frac{d}{2}} \tilde{q}^{\frac{d-1}{8}}}\sum_{a,b,c}C_{abc}^{(0)}v_{a}v_{b}v_{c}\tilde{q}^{n_a+n_b+n_c}\bigg\{1-\frac{l_s^2}{6\pi R^2}\bigg[ \frac{4d^2-24d+31}{4\ln\tilde{q}}\hfill \\
    +\frac{(2d-3)}{8}\big(d-1-8(n_a+n_b+n_c)\big)+\bigg(\frac{(d-1)^2}{64}-\frac{d-1}{4}(n_a+n_b+n_c)\\
   +6\left(n_a^2+n_b^2+n_c^2\right)-\left(n_a+n_b+n_c\right)^2\bigg) \ln \tilde{q}\bigg]+O\left(\partial^4\right)\bigg\}~.
\end{multlined}
\end{equation}
Notably, the locality of the interaction vertices in \eqref{eq_cubic coupling def} requires the coupling constants $C_{abc}^{(0)}$ to be independent of the confining string length $L\sim \ln{\tilde{q}}$. For the open-closed duality to be consistent, the terms scaling as $(\ln{\tilde{q}})^1$ and $(\ln{\tilde{q}})^{-1}$ in equation \eqref{eq_closed channel partition function 2} must match exactly with those obtained from the open-channel calculation. That is indeed the case. 

To make a direct comparison, we collect the expansion in $\tilde{q}$ of the results \eqref{eq_equilateral baryon}, \eqref{eq_nonlinear 1}, \eqref{eq_nonlinear 2}, \eqref{eq_nonlinear 3}, and \eqref{eq_nonlinear 4} as follows
\begin{equation}
\label{eq_expansion collection 1}
\begin{aligned}
    \mathcal{Z}^{(0)}={}&\left(\frac{\pi R}{L}\right)^{d-\frac{3}{2}}\frac{1}{2^{\frac{d}{2}}\tilde{q}^{\frac{d-1}{8}}}\left(1+(d-3) \tilde{q}+O\left(\tilde{q}^2\right)\right)~;\\
    \langle S_\text{junction}^{(1)}\rangle={}&-\frac{(2+d) M l_s^2}{36 R}\left(1+24 \tilde{q}+O\left(\tilde{q}^2\right)\right)~;\\
    \langle S_\text{strings}^{(2)}\rangle={}&\frac{l_s^2}{\pi R^2}\biggl[\left(\frac{4 d^2-20 d+23}{24 (\ln \tilde{q})}+\frac{6 d^2-25 d+19}{144}+\frac{(d-1)^2}{384}(\ln \tilde{q})\right)\\
    &~~~~+\left(-\frac{2 d^2-15 d+27}{6}-\frac{(d^2-16 d+39)}{24}(\ln \tilde{q})\right) \tilde{q}+O\left(\tilde{q}^2\right)\biggr]~;\\
    \langle(S_\text{junction}^{(1)})^2\rangle={}&\frac{M^2 l_s^4}{R^2}\left(\frac{(d+2)^2}{1296}+\frac{d^2-8 d-68}{27} \tilde{q}+O\left(\tilde{q}^2\right)\right)~;\\
    \langle(S_\text{displacement}^{(1)})^2\rangle={}&\frac{l_s^2}{\pi R^2}\left(\frac{d-2}{3(\ln \tilde{q})}+\frac{d-1}{18}-\frac{2(d-3)}{3}(2+\ln \tilde{q}) \tilde{q}+O\left(\tilde{q}^2\right)\right)~.
\end{aligned}
\end{equation}
Matching the closed-channel scattering amplitudes \eqref{eq_closed channel partition function 2} with the open-channel partition function \eqref{eq_expansion collection 1}, we obtain the following two cubic coupling constants:
\begin{equation}
\label{eq_coupling const 2}
\mathtoolsset{multlined-width=0.9\displaywidth}
\begin{multlined}
C^{(0)}_{\textbf{0}\textbf{0}\textbf{0}}=e^{-2 \pi R M}\left[1+\frac{(d+2) M l_s^2}{36 R}+\frac{5(d-1)l_s^2}{144\pi R^2}+\frac{(d+2)^2 M^2l_s^4}{2592R^2}+O\left(R^{-3}\right)\right]~,\hfill\\
C^{(0)}_{\textbf{0}\textbf{0}\textbf{1}}=\frac{e^{-2 \pi R M}}{3\sqrt{d-1}}\left[(d-3)+\frac{(d+2)(d+21)M l_s^2}{36 R}
+\frac{(d-3)(5d-173)l_s^2}{144\pi R^2}\right.\hfill\\
\left.{}+\frac{(d^3+49d^2-392d-3276)M^2l_s^4}{2592R^2}+O\left(R^{-3}\right)\right]~,
\end{multlined}
\end{equation}
which are independent of $L\sim \ln{\tilde{q}}$. In particular, we note that $C^{(0)}_{\textbf{0}\textbf{0}\textbf{1}}=O(R^{-3})$ when $d=3$ and $M=0$. This agrees with the selection rule \eqref{eq_selection rule} that follows from the $\mathbb{Z}_2$ symmetry \eqref{eq_duality defect} of the low-energy effective theory.

\section{Ground state energies of probe baryons}
\label{sec_ground state energies of probe baryons}

In this section, we use the effective field theory \eqref{eq_baryon action} to make predictions for baryons that can be tested in lattice simulations. In particular, we focus on the ground state energies of the probe baryons shown in figure \ref{pic_baryon config}. When the baryon is treated as a point-like particle at long distances, the ground state energy is also referred to as the static mass of the baryon. See, e.g.,  \cite{Takahashi:2000te, Takahashi:2002bw, Jahn:2003uz, Koma:2017hcm, Caselle:2025elf} and references therein for numerical studies on this subject.

The ground state energy $E_{\text{GS}}$ is dominated by the classical contributions from the effective actions \eqref{eq_baryon strings order -2} and \eqref{eq_baryon junction order -1}. We find that
\begin{equation}
\label{eq_gs energy order}
    E_{\text{GS}}=\frac{L_x+L_y+L_z}{l_s^2}+M+E^{(0)}_{\text{GS}}+O\left(\partial^2\right)~,
\end{equation}
where $E^{(0)}_{\text{GS}}$ denotes the leading correction from quantum fluctuations. Since the order 0 effective action is entirely fixed by $S_{\text{strings}}^{(0)}$ in equation \eqref{eq_baryon strings order 0}, the leading correction $E^{(0)}_{\text{GS}}=E^{(0)}_{\text{GS}}(L_x,L_y,L_z)$ is independent of both the string tension $l_s^{-2}$ and the junction mass $M$. While a closed form for $E^{(0)}_{\text{GS}}$ is not readily available, we note that it can be obtained from the open channel partition function \eqref{eq_partition function order 0 general} when the ratios of $L_x$, $L_y$, $L_z$ are rational. We refer to equation \eqref{eq_app_gs energy formula} and the discussion in Appendix \ref{sec_app_rational points} for further details. A few results of equation \eqref{eq_app_gs energy formula} are listed in table \ref{tab_special energy}.
{\renewcommand{\arraystretch}{1.3}
\begin{table}[h!]
\centering
\begin{tabular}{>{\centering\arraybackslash}p{4cm}|>{\centering\arraybackslash}p{6cm}}
  $L_x:L_y:L_z$ & $E^{(0)}_{\text{GS}}\times \min(L_x,L_y,L_z)$\\
  \hline
    $1:1:2$ & $\approx -0.159d+0.323$\\
    $1:1:3$ & $\approx-0.144d+0.298$\\
    $1:3:6$ & $\approx-0.083d+0.179$\\
    $2:3:5$ & $\approx-0.130d+0.265$\\
\end{tabular}
\caption{Leading quantum corrections to the ground-state energy of probe baryons. In this table, we present $E^{(0)}_{\text{GS}}$ for probe baryons at rational points (see Appendix \ref{sec_app_rational points}). The coefficients are displayed to three decimal places.\label{tab_special energy}}
\end{table}}

\begin{figure}[thb]
\centering
\includegraphics[width=\textwidth]{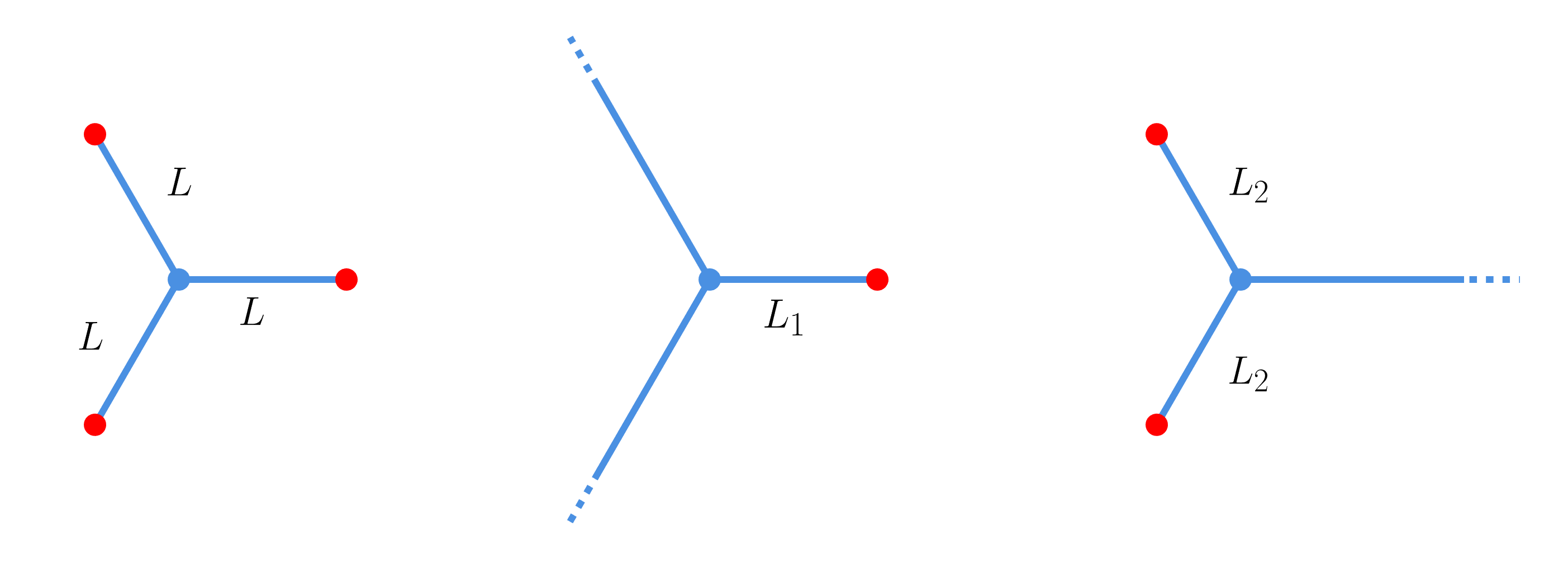}
  \caption{Different limits of the probe baryon configuration. Left: the equilateral baryon; Middle: the isosceles baryon with $L_1\ll L_2$; Right: the isosceles baryon with $L_1\gg L_2$. In the middle and right panels, the long strings are taken to extend to spatial infinity.} \label{pic_baryon special config}
\end{figure}

We now turn to special baryon configurations, starting with
\begin{equation}
    \text{isosceles baryon}~:~~L_x=L_1~,~L_y=L_z=L_2~.
\end{equation}
In particular, we consider the two limits $L_1\ll L_2$ and $L_1\gg L_2$. See also figure \ref{pic_baryon special config}. In either limit, the quantum fluctuations on the short and the long strings are associated with two widely separated scales. For isosceles baryons with $L_1\ll L_2$, we find that
\begin{equation}
\label{eq_isosceles L1<<L2}
    E_{\text{GS}}^{(0)}=-\frac{(d-2)
   \text{Li}_2\left(\frac{1}{3}\right)+\text{Li}_2\left
   (-\frac{1}{3}\right)}{4 \pi  L_1}+O\left(L_2^{-1}\right)~.
\end{equation}
We interpret this as the correction to the ground state energy of a single confining string stretching between a static quark and the baryon junction. On the other hand, we find that for isosceles baryons with $L_1\gg L_2$:
\begin{equation}
\label{eq_isosceles L1>>L2}
    E_{\text{GS}}^{(0)}=-\frac{(2d-5)\pi}{48L_2}-\frac{(d-2)
   \text{Li}_2\left(-\frac{1}{3}\right)+\text{Li}_2\left(
   \frac{1}{3}\right)}{4 \pi  L_2}+O\left(L_1^{-1}\right)~.
\end{equation}

For the equilateral baryons in equation \eqref{eq_equilateral def}, the ground-state energy can be analyzed more explicitly. This energy can be computed using the open-channel representation of the partition function \eqref{eq_perturbative expansion}, with contributions given in equations \eqref{eq_equilateral baryon}, \eqref{eq_nonlinear 1}, \eqref{eq_nonlinear 2}, \eqref{eq_nonlinear 3}, and \eqref{eq_nonlinear 4}. We organize these contributions in an expansion of the modular parameter $q=\exp(-\pi\beta/L)$ as follows:
\begin{equation}
\label{eq_expansion collection 2}
\begin{aligned}
    \mathcal{Z}^{(0)}={}&q^{\frac{2-d}{16}} \Big[1+d\sqrt{q}+O\left(q\right)\Big]~;\\
    \langle S_\text{junction}^{(1)}\rangle={}&\frac{Ml_s^2}{L}(\ln{q}) \Big[\frac{d+2}{144}+\frac{d+2}{6}\sqrt{q}+O\left(q\right) \Big]~;\\
    \langle S_\text{strings}^{(2)}\rangle={}&\frac{l_s^2}{L^2}(\ln{q})\Big[\frac{\pi  (d-2)^2}{1536}-\frac{\pi(d^2-6d-8)}{96} \sqrt{q}+O\left(q\right)\Big]~;\\
    \langle(S_\text{junction}^{(1)})^2\rangle={}&\frac{M^2l_s^4}{L^2}(\ln{q})\Big[\frac{d+6}{216}+\frac{d+6}{9} \sqrt{q}+O\left(q\right)\Big]\\
    {}&+\frac{M^2l_s^4}{L^2}(\ln{q})^2\Big[\frac{(d+2)^2}{20736}+\frac{(d^2+16 d+76)}{432}  \sqrt{q}+O\left(q\right)\Big]~;\\
    \langle(S_\text{displacement}^{(1)})^2\rangle={}&\frac{l_s^2}{L^2}(\ln{q})\Bigl[-\frac{\pi }{6}\sqrt{q}+O\left(q\right)\Bigr]~.
\end{aligned}
\end{equation}
Comparing the leading terms in the expansion \eqref{eq_expansion collection 2} with the Boltzmann weight $\exp{(-\beta E_{\text{GS}})}$ yields the ground state energy:
\begin{equation}
\label{eq_gs energy to order 2}
\mathtoolsset{multlined-width=0.9\displaywidth}
\begin{multlined}
E_{\text{GS}}=\frac{3L}{l_s^2}+M-\frac{(d-2)\pi}{16L}-\frac{ (d+2)\pi  M l_s^2}{144 L^2}\hfill\\
\hfill+\frac{(d+6)\pi M^2l_s^2}{432L^3}-\frac{(d-2)^2\pi^2l_s^2}{1536L^3}+O\left(L^{-4}\right)~.
\end{multlined}
\end{equation}
Notably, the quantum corrections in \eqref{eq_gs energy to order 2} are fixed by the two classical parameters (i.e., the string tension $l_s^{-2}$ and the junction mass $M$) up to the next-to-next-to-leading order. This is a consequence of the Poincar\'e symmetry and diffeomorphism invariance on the confining string worldsheet. In lattice simulations \cite{Caselle:2025elf}, the junction mass $M$ is extracted by fitting the $L^{-2}$ term in \eqref{eq_gs energy to order 2}. The new $L^{-3}$ result should improve fit precision and deepen our understanding of confinement.

\acknowledgments

We are especially grateful to Zohar Komargodski for many stimulating and helpful discussions on this project. We thank Arkya Chatterjee, Gabriel Cuomo, Leonardo Rastelli, Shu-Heng Shao, and Yunqin Zheng for interesting discussions. We also thank Jan Albert and Zohar Komargodski for thoughtful comments on the draft. XL and SZ are supported in part by the Simons Foundation grant 488657 (Simons Collaboration on the Non-Perturbative Bootstrap), the BSF grant no. 2018204 and NSF award number 2310283. The authors of this paper are ordered alphabetically.

\appendix

\section{Modular functions}
\label{sec_app_modular function}

In this appendix, we collect definitions and properties of modular functions that we have used in the main text. The Dedekind eta function is defined by
\begin{equation}
\label{eq_app_eta}
\eta(q)\equiv q^{\frac{1}{24}}\prod_{n\in\mathbb{N}^+}(1-q^n)~.
\end{equation}
The Eisenstein series is defined by
\begin{equation}
\label{eq_app_Eisenstein}
E_{2k}(q)\equiv 1+\frac{2}{\zeta (1-2k)}\sum_{n\in \mathbb{N}^+}\frac{n^{2k-1}q^n}{1-q^n}~,
\end{equation}
where $\zeta(s)$ is the Riemann zeta function. Finally, the infinite product representation of the odd Jacobi theta function is as follows:
\begin{equation}
\label{eq_app_theta}
    \vartheta_{1}(u;q)\equiv 2q^{\frac{1}{8}}\sin(\pi u)\prod_{n\in\mathbb{N}^+}(1-q^{n})(e^{2\pi i u}-q^{n})(e^{-2\pi i u}-q^{n})~.
\end{equation}

Throughout this paper, we adopt the convention that the modular parameter $q=e^{2\pi i \tau}$ and the dual parameter $\tilde{q}=e^{2\pi i \tilde{\tau}}$, with $\tilde{\tau}=-1/\tau$. The functions \eqref{eq_app_eta} and \eqref{eq_app_theta} are known to satisfy the following modular identities:
\begin{equation}
\label{eq_app_modular trans eta theta}
\begin{aligned}
\eta(q)={}&\sqrt{-i\tilde{\tau}}\eta(\tilde{q})~,\\
\vartheta_{1}(u;q)={} &{-}i\sqrt{-i\tilde{\tau}}\exp{(\pi iu^2 \tilde{\tau})}\vartheta_{11}(u\tilde{\tau};\tilde{q})~.
    \end{aligned}
\end{equation}
As far as the discussion in the main text is concerned, we also note the following two identities of Eisenstein series $E_2$ and $E_4$:
\begin{equation}
\label{eq_app_modular trans Eisenstein}
\begin{aligned}
E_2(q)={}&{-}\frac{6i}{\pi }\tilde{\tau}+\tilde{\tau}^2E_2(\tilde{q})~,\\
E_4(q)={} &\tilde{\tau}^4E_4(\tilde{q})~.\\
    \end{aligned}
\end{equation}

\section{Probe baryons at rational points}
\label{sec_app_rational points}

In this appendix, we discuss special configurations of the probe baryon in figure \ref{pic_baryon config}. We focus on the cases where the ratios of the three confining string lengths are rational:
\begin{equation}
\begin{aligned}
    \text{rational points}~:&~~L_x=N_xL~,~~L_y=N_yL~,~~L_z=N_zL,\\
    &~~\text{s.t.}~~N_x,N_y,N_z\in\mathbb{N}^+~~\text{and}~~\text{gcd}(N_x,N_y,N_z)=1~.
\end{aligned}
\end{equation}
Let us first note several useful mathematical facts concerning $N_x$, $N_y$, and $N_z$. In particular, consider the following two polynomials:
\begin{equation}
\label{eq_app_rational points}
\begin{aligned}
    P_1(r)\equiv{}& 1+\frac{1}{3}(r^{N_x}+r^{N_y}+r^{N_z})-\frac{1}{3}(r^{N_x+N_y}+r^{N_y+N_z}+r^{N_z+N_x})-r^{N_x+N_y+N_z}~,\\
    P_2(r)\equiv{}& 1-\frac{1}{3}(r^{N_x}+r^{N_y}+r^{N_z})-\frac{1}{3}(r^{N_x+N_y}+r^{N_y+N_z}+r^{N_z+N_x})+r^{N_x+N_y+N_z}~.
\end{aligned}
\end{equation}
Both $P_1(r)$ and $P_2(r)$ have all of their roots on the unit circle $|r|=1$ \cite{lakatos2004self}. We can therefore write the factorization
\begin{equation}
    P_1(r)=(1-r)\prod_{u}\big(e^{2\pi i u}-r\big)~,~~
    P_2(r)=(1-r)^2\prod_{v}\big(e^{2\pi i v}-r\big)~,
\end{equation}
where $0<u<1$ denotes the phases of the $(N_x+N_y+N_z-1)$ non-identity roots of $P_1$, and $0<v<1$ denotes the phases of the $(N_x+N_y+N_z-2)$ non-identity roots of $P_2$. These phases are subject to the following conditions
\begin{equation}
\begin{aligned}
    \prod_u \big(2\sin(\pi u)\big)=&\frac{4}{3}(N_x+N_y+N_z)~,\\
    \prod_v \big(2\sin(\pi v)\big)=&\frac{2}{3}(N_xN_y+N_yN_z+N_zN_x)~.
\end{aligned}
\end{equation}

We now turn to the baryon partition function \eqref{eq_partition function order 0 general} at the rational points \eqref{eq_app_rational points}. Adopting the notation $q=\exp{(-2\pi^2R/L)}$ and $\tilde{q}=\exp{(-2L/R)}$ as in the previous sections, we find that
\begin{equation}
\label{eq_app_inf prod 1}
\begin{aligned}
       &\prod_{n\in\mathbb{N}^+}\frac{\tanh{(\frac{nN_xL}{R})}+\tanh{(\frac{nN_yL}{R})}+\tanh{(\frac{nN_zL}{R})}}{\tanh{(\frac{nN_xL}{R})}\tanh{(\frac{nN_yL}{R})}\tanh{(\frac{nN_zL}{R})}}\\
       ={}&\prod_{n\in \mathbb{N}^+}\frac{3P_1(\tilde{q}^n)}{ (1-\tilde{q}^{nN_x})(1-\tilde{q}^{nN_y})(1-\tilde{q}^{nN_z})}\\
       ={}&\frac{L}{2\pi R}\sqrt{\frac{N_xN_yN_z}{N_x+N_y+N_z}}\frac{\eta(q)}{\eta(q^{\frac{1}{N_x}})\eta(q^{\frac{1}{N_y}})\eta(q^{\frac{1}{N_z}})}\Big(\prod_uq^{\frac{(u-\frac{1}{2})^2}{4}-\frac{1}{48}}(q^u;q)_\infty\Big)~,
\end{aligned}
\end{equation}
where we have used the $q$-Pochhammer symbol. Similarly,
\begin{equation}
\label{eq_app_inf prod 2}
\begin{aligned}
&\prod_{n\in\mathbb{N}^+}\Big(\coth{(\frac{nN_xL}{R})}+\coth{(\frac{nN_yL}{R})}+\coth{(\frac{nN_zL}{R})}\Big)\\
       ={}&\prod_{n\in \mathbb{N}^+}\frac{3P_2(\tilde{q}^n)}{ (1-\tilde{q}^{nN_x})(1-\tilde{q}^{nN_y})(1-\tilde{q}^{nN_z})}\\
       ={}&\frac{1}{\sqrt{2\pi R(N_x^{-1}+N_y^{-1}+N_z^{-1})}}\frac{\big(\eta(q)\big)^2}{\eta(q^{\frac{1}{N_x}})\eta(q^{\frac{1}{N_y}})\eta(q^{\frac{1}{N_z}})}\Big(\prod_vq^{\frac{(v-\frac{1}{2})^2}{4}-\frac{1}{48}}(q^v;q)_\infty\Big)~.
\end{aligned}
\end{equation}
In equations \eqref{eq_app_inf prod 1} and \eqref{eq_app_inf prod 2}, we have used the fact that both $P_1$ and $P_2$ are real polynomials, as well as the modular identity that follows from equation \eqref{eq_app_modular trans eta theta}:
\begin{equation}
    \tilde{q}^{\frac{1}{12}}\prod_{n\in\mathbb{N}^+}(e^{2\pi i \alpha}-\tilde{q}^{n})(e^{-2\pi i \alpha}-\tilde{q}^{n})=\frac{q^{\frac{1}{2}(\alpha-\frac{1}{2})^2-\frac{1}{24}}}{2\sin(\pi \alpha)}(q^\alpha;q)_\infty(q^{1-\alpha};q)_\infty~.
\end{equation}
In terms of the roots of $P_1$ and $P_2$, we find that the baryon partition function \eqref{eq_partition function order 0 general} takes the form
\begin{equation}
\label{eq_app_rational baryon partition function}
    \mathcal{Z}^{(0)}=\frac{1}{\big(\eta(q)\big)^{2d-3}}\Bigg(\prod_u \frac{q^{\frac{1}{48}-\frac{(u-\frac{1}{2})^2}{4}}}{(q^u;q)_{\infty}}\Bigg)\Bigg(\prod_v \frac{q^{\frac{1}{48}-\frac{(v-\frac{1}{2})^2}{4}}}{(q^v;q)_{\infty}}\Bigg)^{d-2}~.
\end{equation}
This partition function \eqref{eq_app_rational baryon partition function}  indicates that the probe baryon system in the open channel \eqref{eq_baryon open channel} has a unique ground state. Moreover, we find the leading quantum correction to the ground state energy \eqref{eq_gs energy order}:
\begin{equation}
\label{eq_app_gs energy formula}
    E^{(0)}_{\text{GS}}=-\frac{(2d-3)\pi}{24L}-\frac{\pi}{4L}\sum_{u}\Big((u-\frac{1}{2})^2-\frac{1}{12}\Big)-\frac{(d-2)\pi}{4L}\sum_{v}\Big((v-\frac{1}{2})^2-\frac{1}{12}\Big)~.
\end{equation}

Next, we analyze various limits of the probe baryon configurations introduced in sections \ref{sec_Open-closed duality of baryon junctions} and \ref{sec_ground state energies of probe baryons}. For the equilateral baryon with $N_x=N_y=N_z=1$, we obtain
\begin{equation}
    \text{equilateral baryon}~:~~u_1=u_2=v_1=\frac{1}{2}~,
\end{equation}
in which case equation \eqref{eq_app_rational baryon partition function} is reduced to equation \eqref{eq_equilateral baryon}.

For isosceles baryons, we first consider $N_x=1$ and $N_y=N_z=N\gg 1$, corresponding to the  $L_1\ll L_2$ baryon configuration in equation \eqref{eq_isosceles L1<<L2}. See also figure \ref{pic_baryon special config}. In this case, both polynomials $P_1$ and $P_2$ admit two branches of nontrivial roots, with the first exact branch given by
\begin{equation}
\label{eq_app_roots 1}
\begin{aligned}
    \text{isosceles baryons $L_1\ll L_2$}~:~~&u_k^{(1)}=\frac{1}{N}\big(k-\frac{1}{2}\big)~,~~\text{for}~~1\leq k\leq N~;\\
    & v^{(1)}_k=\frac{k}{N}~,~~\text{for}~~1\leq k\leq N-1~.
\end{aligned}
\end{equation}
The remaining roots of $P_1$ and $P_2$ are algebraic, but they cannot be written in closed form for generic $N$. We can solve them perturbatively in the limit $N\gg 1$:
\begin{equation}
\label{eq_app_roots 2}
\mathtoolsset{multlined-width=0.9\displaywidth}
\begin{multlined}
\text{isosceles baryons $L_1\ll L_2$}~:~~\hfill\\
\hfill u_k^{(2)}=\frac{k}{N+1} +\frac{1}{N+1}f^+\big(\frac{k}{N+1}\big)+O\left(N^{-2}\right)~,~~\text{for}~~1\leq k\leq N~;\\
\hfill v^{(2)}_k=\frac{k}{N+1} +\frac{1}{N+1}f^{-}\big(\frac{k}{N+1}\big)+O\left(N^{-2}\right)~,~~\text{for}~~1\leq k\leq N~,
\end{multlined}
\end{equation}
so that these roots are almost uniformly distributed on the unit circle. In equation \eqref{eq_app_roots 2}, we have introduced the function 
\begin{equation}
    f^{\pm}(\alpha)\equiv \alpha+\frac{1}{2\pi}\arg{\Big(\frac{3\pm e^{2\pi i \alpha}}{3e^{2\pi i \alpha }\pm 1}\Big)}~,~~\text{where}~~0\leq \arg <2\pi ~.
\end{equation}
Using the roots in equations \eqref{eq_app_roots 1} and \eqref{eq_app_roots 2}, we find the ground state energy \eqref{eq_app_gs energy formula}
\begin{equation}
     E^{(0)}_{\text{GS}}=-\frac{\pi}{4L}\int_0^1 d\alpha (2\alpha-1)\big(f^+(\alpha)+(d-2)f^-(\alpha)\big)+O\left(N^{-1}\right)~,
\end{equation}
which evaluates to equation \eqref{eq_isosceles L1<<L2}.

We next consider $N_x=N\gg 1$ and $N_y=N_z=1$, corresponding to the  $L_1\gg L_2$ baryon configuration in equation \eqref{eq_isosceles L1>>L2}. See also figure \ref{pic_baryon special config}. The non-trivial roots of the polynomial $P_1$ in this case take the form:
\begin{equation}
\mathtoolsset{multlined-width=0.9\displaywidth}
\begin{multlined}
\text{isosceles baryons $L_1\gg L_2$}~:~~u^{(1)}=\frac{1}{2}~;\hfill\\
\hfill u^{(2)}_k=\frac{k}{N+1} +\frac{1}{N+1}f^{-}\big(\frac{k}{N+1}\big)+O\left(N^{-2}\right)~,~~\text{for}~~1\leq k\leq N~.
\end{multlined}
\end{equation}
On the other hand, the corresponding roots of the polynomial $P_2$ are given by
\begin{equation}
\mathtoolsset{multlined-width=0.9\displaywidth}
\begin{multlined}
\text{isosceles baryons $L_1\gg L_2$}~:\hfill\\
\hfill v_k=\frac{k}{N+1} +\frac{1}{N+1}f^{+}\big(\frac{k}{N+1}\big)+O\left(N^{-2}\right)~,~~\text{for}~~1\leq k\leq N~.
\end{multlined}
\end{equation}
We thus find that the ground state energy \eqref{eq_app_gs energy formula} yields equation \eqref{eq_isosceles L1>>L2}:
\begin{equation}
     E^{(0)}_{\text{GS}}=-\frac{(2d-5)\pi}{48L}-\frac{\pi}{4L}\int_0^1 d\alpha (2\alpha-1)\big((d-2)f^+(\alpha)+f^-(\alpha)\big)+O\left(N^{-1}\right)~.
\end{equation}

\section{Green's functions on a cylinder}
\label{sec_app_greens functions}

This appendix collects the technical details of Green’s functions and zeta-function regularization used in sections \ref{sec_Partition function of the probe baryon} and \ref{sec_regularized partition function}. 

We consider a Euclidean cylinder with coordinates $(\tau,\sigma)$, where $\tau \sim \tau+2\pi R$ and $0\leq \sigma\leq L$. The Green's functions on the cylinder are defined as the solutions to the Poisson equation:
\begin{equation}
    (-\partial_{\tau_1}^2-\partial_{\sigma_1}^2)G(\sigma_1,\sigma_2,\tau_{12})=\delta(\tau_{12})\delta(\sigma_{12})~,
\end{equation}
where $\tau_{ab}\equiv \tau_a-\tau_b$ and $\sigma_{ab}\equiv\sigma_a-\sigma_b$. We denote the Green's function with Dirichlet boundary conditions at $\sigma=0$ and $\sigma=L$ by
\begin{equation}
\label{eq_app_DD green function def}
   G(\sigma_1,\sigma_2,\tau_{12};q)=\frac{1}{\pi R L}\sum_{n\in \mathbb{N}^+}\sum_{m \in \mathbb {Z}}\frac{\sin\left(\frac{n\pi\sigma_1}{L}\right)\sin\left(\frac{n\pi\sigma_2}{L}\right)e^{i\frac{m}{R}\tau_{12}}}{\frac{\pi^2n^2}{L^2}+\frac{m^2}{R^2}}~,
\end{equation}
where $q=\exp{(-2\pi^2R/L)}$ is the modular parameter. On the other hand, the Green's function with Neumann boundary condition at $\sigma=0$ and Dirichlet boundary condition at $\sigma=L$ takes the form
\begin{equation}
\label{eq_app_ND green function def}
    \tilde{G}(\sigma_1, \sigma_2, \tau_{12};q)=\frac{1}{\pi RL} \sum_{r \in \mathbb{N}+\frac{1}{2}} \sum_{m \in \mathbb{Z}} \frac{\cos \left(\frac{r \pi \sigma_1}{L}\right) \cos \bigl(\frac{r \pi \sigma_2}{L}\bigr) e^{i\frac{m}{R}\tau_{12}}}{\frac{\pi^2 r^2}{L^2}+\frac{m^2}{R^2}}~.
\end{equation}

In section \ref{sec_Partition function of the probe baryon}, we use the Dirichlet-Dirichlet Green’s function \eqref{eq_app_DD green function def} to compute the saddle-point action in equation \eqref{eq_order 0 factorize}. It is convenient to define 
\begin{equation}
\label{eq_app_K def}
    K(\tau;q)\equiv \frac{R}{L}+2\sum_{n \in \mathbb{N}^+}n \cos{(\frac{n\tau}{R})}\coth{(\frac{nL}{R})}~,
\end{equation}
such that the saddle point action $S^{(0)}_{\text{strings}}(x_i^\star,y_i^\star, z_i^\star)$ can be written compactly as:
\begin{equation}
\label{eq_app_trapping potential}
\mathtoolsset{multlined-width=0.9\displaywidth}
\begin{multlined}
    S^{(0)}_{\text{strings}}(x_i^\star,y_i^\star, z_i^\star)=-\int \frac{d\tau_1 d\tau_2}{4\pi R^2}\Bigg[\frac{3}{4}\Big( K(\tau_{12};q_y)+K(\tau_{12};q_z)\Big)w_1(\tau_1)w_1(\tau_2)\hfill\\
        +\frac{\sqrt{3}}{2}\Big( K(\tau_{12};q_y)-K(\tau_{12};q_z)\Big)w_1(\tau_1)w_2(\tau_2)\\
    +\frac{1}{4}\Big( 4 K(\tau_{12};q_x)+K(\tau_{12};q_y)+K(\tau_{12};q_z)\Big)w_2(\tau_1)w_2(\tau_2)\\
\hfill+\big(K(\tau_{12};q_x)+K(\tau_{12};q_y)+K(\tau_{12};q_z)\big)w_j(\tau_1)w_j(\tau_2)\Bigg]~.
\end{multlined}
\end{equation}
Using the Gaussian integral determinant of \eqref{eq_app_trapping potential}, we find the partition function \eqref{eq_partition function order 0 general}.

In the following, we focus on the calculations of nonlinear corrections presented in section \ref{sec_regularized partition function}.

\subsection{Wick contraction}

For the equilateral baryon configuration \eqref{eq_equilateral def}, the free theory correlation functions between the NGB fields $x_i$, $y_i$, and $z_i$ are given by linear combinations of the Green's functions \eqref{eq_app_DD green function def} and \eqref{eq_app_ND green function def}. In particular, we find the propagators of the quantum fluctuations in the confining string plane:
\begin{equation}
\label{eq_app_propagators 1}
\mathtoolsset{multlined-width=0.9\displaywidth}
\begin{multlined}
 \langle x_{2}(\tau_1,\sigma_1)x_{2}(\tau_2,\sigma_2)\rangle=\langle y_{2}(\tau_1,\sigma_2)y_{2}(\tau_2,\sigma_2)\rangle=\langle z_{2}(\tau_1,\sigma_2)z_{2}(\tau_2,\sigma_2)\rangle\hfill\\
 \hfill =\frac{1}{3}G(\sigma_1,\sigma_2,\tau_{12};q)+\frac{2}{3}\tilde{G}(\sigma_1,\sigma_2,\tau_{12};q)~,\\
 \langle x_{2}(\tau_1,\sigma_1)y_{2}(\tau_2,\sigma_2)\rangle=\langle y_{2}(\tau_1,\sigma_2)z_{2}(\tau_2,\sigma_2)\rangle=\langle z_{2}(\tau_1,\sigma_2)x_{2}(\tau_2,\sigma_2)\rangle\hfill\\
        \hfill=\frac{1}{3}G(\sigma_1,\sigma_2,\tau_{12};q)-\frac{1}{3}\tilde{G}(\sigma_1,\sigma_2,\tau_{12};q)~.\\
\end{multlined}
\end{equation}
Similarly, we also find
\begin{equation}
\label{eq_app_propagators 2}
\mathtoolsset{multlined-width=0.9\displaywidth}
\begin{multlined}
\langle x_{j}(\tau_1,\sigma_1)x_{k }(\tau_2,\sigma_2)\rangle=\langle y_{j}(\tau_1,\sigma_1)y_{k }(\tau_2,\sigma_2)\rangle=\langle z_{j}(\tau_1,\sigma_1)z_{k }(\tau_2,\sigma_2)\rangle\hfill\\
        \hfill=\frac{1}{3}\delta_{jk}\tilde{G}(\sigma_1,\sigma_2,\tau_{12};q)+\frac{2}{3}\delta_{jk}G(\sigma_1,\sigma_2,\tau_{12};q)~,\\
\langle x_{j}(\tau_1,\sigma_1)y_{k }(\tau_2,\sigma_2)\rangle=\langle y_{j}(\tau_1,\sigma_1)z_{k }(\tau_2,\sigma_2)\rangle=\langle z_{j}(\tau_1,\sigma_1)x_{k }(\tau_2,\sigma_2)\rangle\hfill\\
        \hfill=\frac{1}{3}\delta_{jk}\tilde{G}(\sigma_1,\sigma_2,\tau_{12};q)-\frac{1}{3}\delta_{jk}G(\sigma_1,\sigma_2,\tau_{12};q)~,\\        
\end{multlined}
\end{equation}
where $3\leq j,k\leq d$ denote spatial directions perpendicular to the confining string plane. We note that the other propagators (e.g. $\langle x_2 y_j\rangle$) are identically zero.

We now use Wick's theorem to obtain the nonlinear corrections $\langle S_\text{displacement}^{(2)}\rangle$, $\langle S_\text{strings}^{(2)}\rangle$, $\langle(S_\text{junction}^{(1)})^2\rangle$, and $\langle(S_\text{displacement}^{(1)})^2\rangle$ for the equilateral baryon configuration. First, we find that the quartic term at the baryon junction arising from longitudinal displacements takes the form
\begin{equation}
\label{eq_app_nonlinear 1}
\mathtoolsset{multlined-width=0.9\displaywidth}
\begin{multlined}
    \langle S_\text{displacement}^{(2)}\rangle=\frac{l_s^2}{3} \int d \tau \big[(2d-3)\tilde{G}\left(\partial_{\tau_1}\partial_{\tau_2}\partial_{\sigma_1}G+\partial_{\sigma_1}^2\partial_{\sigma_2}G\right)\hfill\\
\hfill+d\tilde{G}\bigl(\partial_{\tau_1}\partial_{\tau_2}\partial_{\sigma_1}\tilde G+\partial_{\sigma_1}^2\partial_{\sigma_2}\tilde G\bigr)\big]\Big|_{\tau_{12}=0,~\sigma_1=\sigma_2=0}~.
\end{multlined}
\end{equation}
It is straightforward to check that the triple derivatives of the Green's functions \eqref{eq_app_DD green function def} and \eqref{eq_app_ND green function def} at the coincident point $\tau_{12}=0$, $\sigma_1=\sigma_2=0$ satisfy
\begin{equation}
\label{eq_app_triple derivatives}
\partial_{\tau_1}\partial_{\tau_2}\partial_{\sigma_1}G=\partial_{\sigma_1}^2\partial_{\sigma_2}G=\partial_{\tau_1}\partial_{\tau_2}\partial_{\sigma_1}\tilde G=\partial_{\sigma_1}^2\partial_{\sigma_2}\tilde G=0~.
\end{equation}
We therefore conclude $\langle S_\text{displacement}^{(2)}\rangle=0$ as in equation \eqref{eq_nonlinear 1}.

For the worldsheet nonlinear correction $\langle S_\text{strings}^{(2)}\rangle$, we introduce the following short-handed notations
\begin{equation}
\begin{aligned}
    &G_{\upalpha\upbeta}(\sigma)\equiv\big(\partial_{\upalpha_1}\partial_{\upbeta_2}G(\sigma_1,\sigma_2,\tau_{12};q)\big)\Big|_{\tau_{12}=0,~\sigma_1=\sigma_2=\sigma}~,\\
    &\tilde{G}_{\upalpha\upbeta}(\sigma)\equiv\big(\partial_{\upalpha_1}\partial_{\upbeta_2}\tilde{G}(\sigma_1,\sigma_2,\tau_{12};q)\big)\Big|_{\tau_{12}=0,~\sigma_1=\sigma_2=\sigma}~.
\end{aligned}
\end{equation}
These are the double derivatives of the Green's functions in the coincident point limit. In particular, we have $G_{\tau\sigma}(\sigma)=\tilde{G}_{\tau\sigma}(\sigma)=0$. Applying Wick's theorem to $\langle S_\text{strings}^{(2)}\rangle$, we obtain
\begin{equation}
\label{eq_app_nonlinear 2}
\mathtoolsset{multlined-width=0.9\displaywidth}
\begin{multlined}
    \langle S_\text{strings}^{(2)}\rangle=-\frac{l_s^2}{24} \int d \tau d \sigma \Bigl[(4 d^2-4 d-5)\bigl(G_{\sigma \sigma}^2+G_{\tau \tau}^2\bigr)\\
    \phantom{\langle S_\text{strings}^{(2)}\rangle=}-2(4 d^2-20 d+23) G_{\tau \tau} G_{\sigma \sigma}+(d^2+2 d+4)\bigl(\tilde{G}_{\sigma \sigma}^2+\tilde{G}_{\tau \tau}^2\bigr)\hfill \\
    \phantom{\langle S_\text{strings}^{(2)}\rangle=}-2(d^2-2 d-4) \tilde{G}_{\tau \tau} \tilde{G}_{\sigma \sigma}+2(2 d^2+d-4)\bigl(G_{\sigma \sigma} \tilde{G}_{\sigma \sigma}+G_{\tau \tau} \tilde{G}_{\tau \tau}\bigr) \hfill\\
    \hfill-(4 d^2-14 d+8)\bigl(G_{\sigma \sigma} \tilde{G}_{\tau \tau}+G_{\tau \tau} \tilde{G}_{\sigma \sigma}\bigr)\Bigr]~.
\end{multlined}
\end{equation}
The integrals in equation \eqref{eq_app_nonlinear 2} are divergent, and we evaluate them using zeta-function regularization in the following. See equations \eqref{eq_app_int 1}, \eqref{eq_app_int 2}, and \eqref{eq_app_int 3}.

The nonlinear correction arising from the baryon junction kinetic term reads
\begin{equation}
\label{eq_app_nonlinear 3}
\mathtoolsset{multlined-width=0.9\displaywidth}
\begin{multlined}
\langle(S_\text{junction}^{(1)})^2\rangle=\frac{(d+2)^2}{36}M^2l_s^4\int d\tau_1d\tau_2 \big(\tilde{G}_{\tau\tau}^2\big)\Big|_{\sigma=0}\hfill\\
    \hfill +\frac{d+6}{18}M^2l_s^4\int d\tau_1d\tau_2 \big(\partial_{\tau_1}\partial_{\tau_2}\tilde{G}(\tau_{12})\big)^2\Big|_{\sigma_1=\sigma_2=0}~.
\end{multlined}
\end{equation}
We refer the reader to equations \eqref{eq_app_int 4} and \eqref{eq_app_int 5} for the regularization of this divergent integral. Finally, the longitudinal displacements contribute the following nonlinear corrections
\begin{equation}
\label{eq_app_nonlinear 4}
\mathtoolsset{multlined-width=0.9\displaywidth}
\begin{multlined}
    \langle (S_\text{displacement}^{(1)})^2\rangle=\frac{l_s^2}{3} \int d \tau_1 d \tau_2  \big[(d-2)\tilde{G}(\tau_{12})\bigl(\partial_{\sigma_1} \partial_{\sigma_2} G(\tau_{12})\bigr)^2\\
    +\tilde{G}(\tau_{12})\bigl(\partial_{\tau_1} \partial_{\tau_2} \tilde{G}(\tau_{12})\bigr)^2-2\bigl(\partial_{\tau_1}\tilde{G}(\tau_{12})\bigr)^2\partial_{\tau_1} \partial_{\tau_2} \tilde{G}(\tau_{12})\big]\Big|_{\sigma_1=\sigma_2=0}~,
\end{multlined}
\end{equation}
which we regularize in equations \eqref{eq_app_int 6}, \eqref{eq_app_int 7} and \eqref{eq_app_int 8}.

\subsection{Zeta-function regularization}

In this section, we elaborate on the regularization scheme used in section \ref{sec_regularized partition function}. 

For the divergent integrals in equation \eqref{eq_app_nonlinear 2}, we note the following infinite sums that yield Eisenstein series \eqref{eq_app_Eisenstein}
\begin{equation}
\label{eq_app_sum 1}
    \begin{aligned}
        \sum_{n \in \mathbb{N}^{+}} \sum_{m \in \mathbb{Z}} \frac{\frac{\pi^2 n^2}{L^2}}{\frac{\pi^2 n^2}{L^2}+\frac{m^2}{R^2}} &=-\frac{\ln q}{2}\sum_{n\in\mathbb{N}}n\frac{1+q^n}{1-q^n} =\frac{\ln q}{24} E_2(q)~,\\
        \sum_{n \in \mathbb{N}^{+}} \sum_{m \in \mathbb{Z}} \frac{\frac{m^2}{R^2}}{\frac{\pi^2 n^2}{L^2}+\frac{m^2}{R^2}} &=-\sum_{n \in \mathbb{N}^{+}} \sum_{m \in \mathbb{Z}} \frac{\frac{\pi^2 n^2}{L^2}}{\frac{\pi^2 n^2}{L^2}+\frac{m^2}{R^2}} =-\frac{\ln q}{24} E_2(q)~,
    \end{aligned}
\end{equation}
where we have used the zeta-function identities
\begin{equation}
     \sum_{n\in\mathbb{N}}n^1=\zeta(-1)=-\frac{1}{12}~,~~\text{and}~~   \sum_{m\in\mathbb{Z}}m^0=1+2\zeta(0)=0~.
\end{equation}
Similarly, we obtain the following
\begin{equation}
\label{eq_app_sum 2}
\mathtoolsset{multlined-width=0.9\displaywidth}
\begin{multlined}
    \sum_{n \in \mathbb{N}^{+}}\left(\sum_{m \in \mathbb{Z}} \frac{\frac{\pi^2 n^2}{L^2}}{\frac{\pi^2 n^2}{L^2}+\frac{m^2}{R^2}}\right)^2=\sum_{n \in \mathbb{N}^{+}}\left(\sum_{m \in \mathbb{Z}} \frac{\frac{m^2}{R^2}}{\frac{\pi^2 n^2}{L^2}+\frac{m^2}{R^2}}\right)^2\hfill\\
    \hfill =\frac{(\ln q)^2}{4} \sum_{n\in \mathbb{N}} n^2 \frac{\left(1+q^n\right)^2}{\left(1-q^n\right)^2}=(\ln q)^2 \sum_{n\in \mathbb{N}} \frac{n^2 q^n}{\left(1-q^n\right)^2}=\frac{(\ln q)^2}{288}\left[E_4(q)-(E_2(q))^2\right]~,
\end{multlined}
\end{equation}
where we have used the zeta-function identity
\begin{equation}
    \sum_{n\in\mathbb{N}}n^2=\zeta(-2)=0~.
\end{equation}
These infinite sums appear in the integral of the coincident point functions $G_{\tau\tau}$ and $G_{\sigma\sigma}$. With equations \eqref{eq_app_sum 1} and \eqref{eq_app_sum 2}, we find
\begin{equation}
\label{eq_app_int 1}
    \begin{aligned}
        &\int d\tau d\sigma  G_{\tau \tau}^2 =\int d\tau d\sigma  G_{\sigma \sigma}^2 =-\frac{\pi \ln q}{576 L^2} E_4(q)~,\\  
        &\int d\tau d\sigma  G_{\tau \tau} G_{\sigma \sigma} =-\frac{\pi \ln q}{576 L^2}\left[E_4(q)-2(E_2(q))^2\right]~.
    \end{aligned}
\end{equation}

Other infinite sums appearing in the integral \eqref{eq_app_nonlinear 2} include
\begin{equation}
\label{eq_app_sum 3}
\begin{aligned}
\sum_{r\in\mathbb{N}+\frac{1}2{}}\sum_{m\in\mathbb{Z}}\frac{\frac{\pi^2r^2}{L^2}}{\frac{\pi^2r^2}{L^2}+\frac{m^2}{R^2}}&=-\frac{\ln q}{2}\sum_{r\in\mathbb{N}+\frac{1}{2}}r\frac{1+q^r}{1-q^r}=-\frac{\ln q}{48}\bigl(2E_2(q)-E_2(\sqrt{q})\bigr)~,\\
    \sum_{r\in\mathbb{N}+\frac{1}{2}}\sum_{m\in\mathbb{Z}}\frac{\frac{m^2}{R^2}}{\frac{\pi^2r^2}{L^2}+\frac{m^2}{R^2}}&=-\sum_{r\in\mathbb{N}+\frac{1}2{}}\sum_{m\in\mathbb{Z}}\frac{\frac{\pi^2r^2}{L^2}}{\frac{\pi^2r^2}{L^2}+\frac{m^2}{R^2}}=\frac{\ln q}{48}\bigl(2E_2(q)-E_2(\sqrt{q})\bigr)~,
\end{aligned}
\end{equation}
where we have used
\begin{equation}
    \sum_{r\in \mathbb{N}+\frac{1}{2}}r^1=\zeta (-1,\frac{1}{2})=\frac{1}{24}~.
\end{equation}
Analogous to the sum in \eqref{eq_app_sum 2}, we note
\begin{equation}
\label{eq_app_sum 4}
\mathtoolsset{multlined-width=0.9\displaywidth}
\begin{multlined}
    \sum_{r\in \mathbb{N}+\frac{1}{2}}\left(\sum_{m \in \mathbb{Z}} \frac{\frac{\pi^2 r^2}{L^2}}{\frac{\pi^2 r^2}{L^2}+\frac{m^2}{R^2}}\right)^2=\sum_{r\in \mathbb{N}+\frac{1}{2}}\left(\sum_{m \in \mathbb{Z}} \frac{\frac{m^2}{R^2}}{\frac{\pi^2 r^2}{L^2}+\frac{m^2}{R^2}}\right)^2 \hfill\\
=\frac{(\ln q)^2}{4} \sum_{r\in \mathbb{N}+\frac{1}{2}} r^2 \frac{\left(1+q^r\right)^2}{\left(1-q^r\right)^2}=(\ln q)^2 \sum_{r\in \mathbb{N}+\frac{1}{2}} \frac{r^2 q^r}{\left(1-q^r\right)^2}\\
\hfill =\frac{(\ln q)^2}{1152}\left[E_4(\sqrt{q})-4 E_4(q)-(E_2(\sqrt{q}))^2+4(E_2(q))^2\right]~,
\end{multlined}
\end{equation}
where the zeta-function regularization reads
\begin{equation}
    \sum_{r\in \mathbb{N}+\frac{1}{2}}r^2=\zeta (-2,\frac{1}{2})=0~.
\end{equation}
With equations \eqref{eq_app_sum 3} and \eqref{eq_app_sum 4}, we obtain the integrals of the coincident point functions $\tilde{G}_{\tau\tau}$ and $\tilde{G}_{\sigma\sigma}$:
\begin{equation}
\label{eq_app_int 2}
\mathtoolsset{multlined-width=0.9\displaywidth}
\begin{multlined}
        \int d\tau d\sigma \tilde{G}_{\tau \tau}^2 =\int d\tau d\sigma \tilde{G}_{\sigma \sigma}^2 \hfill\\
        \hfill=\frac{\pi \ln q}{2304 L^2}\left[4 E_4(q)+4 E_2(\sqrt{q}) E_2(q)-E_4(\sqrt{q})-8(E_2(q))^2\right]~,\\
        \int d\tau d\sigma \tilde{G}_{\tau \tau}\tilde{G}_{\sigma \sigma} =\frac{\pi \ln q}{2304 L^2}\left[4 E_4(q)+2(E_2(\sqrt{q}))^2-4 E_2(\sqrt{q}) E_2(q)-E_4(\sqrt{q})\right]~.\hfill\\
\end{multlined}
\end{equation}
Finally, we compute the following mixed terms using \eqref{eq_app_sum 1} and \eqref{eq_app_sum 3}:
\begin{equation}
\label{eq_app_int 3}
\mathtoolsset{multlined-width=0.9\displaywidth}
\begin{multlined}
    \int d \tau d \sigma  G_{\sigma \sigma} \tilde{G}_{\sigma \sigma}=\int d \tau d \sigma  G_{\tau \tau} \tilde{G}_{\tau \tau}=-\int d \tau d \sigma  G_{\tau \tau} \tilde{G}_{\sigma \sigma}=-\int d \tau d \sigma  G_{\sigma \sigma} \tilde{G}_{\tau \tau}\hfill\\
    \hfill =\frac{\pi}{1152 L^2} \ln q\left[2(E_2(q))^2-E_2(\sqrt{q}) E_2(q)\right]~.
\end{multlined}
\end{equation}

Next, we turn to the divergent integrals in equation \eqref{eq_app_nonlinear 3}. It follows from the infinite sum \eqref{eq_app_sum 3} that
\begin{equation}
\label{eq_app_int 4}
    \int d\tau_1d\tau_2 \big(\tilde{G}_{\tau\tau}^2\big)\Big|_{\sigma=0}=\Big(\int d \tau \big(\tilde{G}_{\tau\tau}\big)\Big|_{\sigma=0}\Big)^2=\frac{(\ln q)^2}{576L^2}\left(2 E_2(q)-E_2(\sqrt{q})\right)^2~.
\end{equation}
For the integral on the second line of equation \eqref{eq_app_nonlinear 3}, we consider the infinite sum
\begin{equation}
\label{eq_app_sum 5}
\begin{aligned}
&\sum_{r,r' \in \mathbb{N}+\frac{1}{2}} \sum_{m \in \mathbb{Z}} \frac{\frac{m^4}{R^4}}{\left(\frac{\pi^2 r^2}{L^2}+\frac{m^2}{R^2}\right)\left(\frac{\pi^2 r^{\prime 2}}{L^2}+\frac{m^2}{R^2}\right)} \\
={}&\frac{\ln q}{2} \sum_{r,r' \in \mathbb{N}+\frac{1}{2}} \frac{1}{r^2-r^{\prime 2}}\left(r^3 \frac{1+q^r}{1-q^r}-r^{\prime 3} \frac{1+q^{r'}}{1-q^{r'}}\right)\\
={}&\frac{\ln q}{2} \sum_{r\in \mathbb{N}+\frac{1}{2}} r \frac{1+q^r}{1-q^r}+\frac{(\ln q)^2}{2} \sum_{r\in \mathbb{N}+\frac{1}{2}} \frac{r^2 q^r}{\left(1-q^r\right)^2}\\
={}&\frac{\ln q}{48}\bigl(2 E_2(q)-E_2(\sqrt{q})\bigr) +\frac{(\ln q)^2}{2304}\left[E_4(\sqrt{q})-4 E_4(q)-(E_2(\sqrt{q}))^2+4(E_2(q))^2\right]~,
\end{aligned}
\end{equation}
where in the third line we have used the convergent sum
\begin{equation}
    \sum_{r'\in(\mathbb{N}+\frac{1}{2})/\{r\}}\frac{1}{r^{\prime2}-r^2}=\frac{1}{4r^2}~.
\end{equation}
Using equation \eqref{eq_app_sum 5}, we find 
\begin{equation}
\label{eq_app_int 5}
\mathtoolsset{multlined-width=0.9\displaywidth}
\begin{multlined}
    \int d\tau_1d\tau_2 \big(\partial_{\tau_1}\partial_{\tau_2}\tilde{G}(\tau_{12})\big)^2\Big|_{\sigma_1=\sigma_2=0}=\frac{\ln q}{12L^2}\left(2 E_2(q)-E_2(\sqrt{q})\right)\hfill\\
    \hfill +\frac{(\ln q)^2}{576 L^2}\left[E_4(\sqrt{q})-4 E_4(q)-(E_2(\sqrt{q}))^2+4(E_2(q))^2\right]~.
\end{multlined}
\end{equation}

Finally, we consider the integrals in equation \eqref{eq_app_nonlinear 4}. There are three infinite sums involved in these integrals, and we start with
\begin{equation}
\label{eq_app_sum 6}
\begin{aligned}
    &\sum_{m_1,m_2,m_3\in \mathbb{Z}} \sum_{r_1,r_2,r_3 \in \mathbb{N}+\frac{1}{2}} \delta_{m_1+m_2+m_3}\frac{\frac{m_1}{R}}{\frac{\pi^2 r_1^2}{L^2}+\frac{m_1^2}{R^2}} \frac{\frac{m_2}{R}}{\frac{\pi^2 r_2^{2}}{L^2}+\frac{m_2^{2}}{R^2}} \frac{\frac{m_3^2}{R^2}}{\frac{\pi^2 r_3^{2}}{L^2}+\frac{m_3^2}{R^2}} \\
    ={}&{-}\frac{L^2 }{16} \ln\tilde q\sum_{m, m^{\prime} \in \mathbb{Z}}(m+m^{\prime}) \frac{1-\tilde{q}^m}{1+\tilde{q}^m} \frac{1-\tilde{q}^{m^{\prime}}}{1+\tilde{q}^{m^{\prime}}} \frac{1-\tilde{q}^{m+m^{\prime}}}{1+\tilde{q}^{m+m^{\prime}}} \\
    ={}&{-}\frac{L^2 }{8} \ln\tilde q\sum_{m,m'\in\mathbb{Z}}(m+m')\left(\frac{1}{1-\tilde{q}^m}+\frac{1}{1-\tilde{q}^{m^{\prime}}}-\frac{1}{1+\tilde{q}^{m+m^{\prime}}}-\frac{1}{2}\right) =0~.
\end{aligned}
\end{equation}
Using equation \eqref{eq_app_sum 6}, we obtain the integral
\begin{equation}
\label{eq_app_int 6}
    \int d \tau_1 d \tau_2  \big(\bigl(\partial_{\tau_1}\tilde{G}(\tau_{12})\bigr)^2\partial_{\tau_1} \partial_{\tau_2} \tilde{G}(\tau_{12})\big)\Big|_{\sigma_1=\sigma_2=0}=0~.
\end{equation}
We also need the following infinite sum:

\begin{equation}
\label{eq_app_sum 7}
\mathtoolsset{multlined-width=0.9\displaywidth}
\begin{multlined}
\phantom{{}={}}\sum_{m_1,m_2,m_3 \in \mathbb{Z}} \sum_{n, n' \in \mathbb{N}^{+}} \sum_{r \in \mathbb{N}+\frac{1}{2}} \delta_{m_1+m_2+m_3}\frac{\frac{\pi^2 n^2}{L^2}}{\frac{\pi^2 n^2}{L^2}+\frac{m^2}{R^2}} \frac{\frac{\pi^2 n^{\prime2}}{L^2}}{\frac{\pi^2 n^{\prime2}}{L^2}+\frac{m^{\prime2}}{R^2}}\frac{1}{\frac{\pi^2 r^2}{L^2}+\frac{m_3^2}{R^2}}\hfill \\
=-\frac{L^2}{16} \ln\tilde q \sum_{m, m^{\prime} \in \mathbb{Z}} \frac{m m^{\prime}}{m+m^{\prime}} \frac{1+\tilde{q}^m}{1-\tilde{q}^m} \frac{1+\tilde{q}^{m^{\prime}}}{1-\tilde{q}^{m^{\prime}}} \frac{1-\tilde{q}^{m+m^{\prime}}}{1+\tilde{q}^{m+m^{\prime}}}\hfill \\
=\frac{L^2}{16}\ln\tilde q\Biggl[\frac{\ln \tilde{q}}{2} \sum_{m \in \mathbb{Z}} m^2\left(\frac{1+\tilde{q}^m}{1-\tilde{q}^m}\right)^2+\sum_{m \in \mathbb{Z}}m\left(\frac{4}{1-\tilde{q}^m}-1\right)\Biggr]\hfill \\
=-\frac{L^2}{16}\ln\tilde q\Biggl[\frac{2}{\ln \tilde{q}}-4 \ln \tilde{q} \sum_{n \in \mathbb{N}^{+}} \frac{n^2 \tilde{q}^n}{\left(1-\tilde{q}^n\right)^2}+\frac{1}{3}-8 \sum_{n \in \mathbb{N}^{+}} \frac{n \tilde{q}^n}{1-\tilde{q}^n}\Biggr]\hfill\\
=L^2\left[\frac{(\ln \tilde{q})^2}{1152}\left(E_4(\tilde{q})-(E_2(\tilde{q}))^2\right)-\frac{1}{8}-\frac{\ln \tilde{q}}{48} E_2(\tilde{q})\right]=\frac{\pi^4R^2}{288}\left[E_4(q)-(E_2(q))^2\right]~,\hfill\\
\end{multlined}
\end{equation}
where we have used the modular transformation \eqref{eq_app_modular trans Eisenstein}. Equation \eqref{eq_app_sum 7} simplifies the integral
\begin{equation}
\label{eq_app_int 7}
    \int d \tau_1 d \tau_2  \big(\tilde{G}(\tau_{12})\big(\partial_{\sigma_1} \partial_{\sigma_2} G(\tau_{12})\big)^2\big)\Big|_{\sigma_1=\sigma_2=0}=\frac{\pi \ln q}{144 L^2}\left[(E_2(q))^2-E_4(q)\right]~.
\end{equation}
The last infinite sum we need is as follows
\begin{equation}
\label{eq_app_sum 8}
\mathtoolsset{multlined-width=0.9\displaywidth}
\begin{multlined}
    \phantom{{}={}}\sum_{m_1,m_2,m_3 \in \mathbb{Z}} \sum_{r, r', r'' \in \mathbb{N}+\frac{1}{2}}  \delta_{m_1+m_2+m_3}\frac{\frac{m_1^2}{R^2}}{\frac{\pi^2 r_1^2}{L^2}+\frac{m_1^2}{R^2}} \frac{\frac{m_2^{2}}{R^2}}{\frac{\pi^2 r_2^{2}}{L^2}+\frac{m_2^{2}}{R^2}}\frac{1}{\frac{\pi^2 r_3^{2}}{L^2}+\frac{m_3^2}{R^2}}\hfill \\
    = -\frac{L^2}{16}\ln\tilde q \sum_{m, m^{\prime} \in \mathbb{Z}} \frac{m m^{\prime}}{m+m^{\prime}} \frac{1-\tilde{q}^m}{1+\tilde{q}^m} \frac{1-\tilde{q}^{m^{\prime}}}{1+\tilde{q}^{m^{\prime}}} \frac{1-\tilde{q}^{m+m^{\prime}}}{1+\tilde{q}^{m+m^{\prime}}}\hfill \\
    =\frac{L^2}{16}\ln\tilde q\left[\frac{\ln \tilde{q}}{2} \sum_{m \in \mathbb{Z}} m^2\left(\frac{1-\tilde{q}^m}{1+\tilde{q}^m}\right)^2+\sum_{m \in \mathbb{Z}}m\left(\frac{4}{1+\tilde{q}^m}-1\right)\right]\hfill \\
    =-\frac{L^2}{16}\ln\tilde q\Biggl[4 \ln \tilde{q} \sum_{n \in \mathbb{N}^{+}} \frac{n^2 \tilde{q}^n}{(1+\tilde{q}^n)^2}+\frac{1}{3}+8 \sum_{n \in \mathbb{N}^{+}} \frac{n \tilde{q}^n}{1+\tilde{q}^n}\Biggr]\hfill \\
    =\frac{L^2}{48}\ln\tilde q\bigl(E_2(\tilde{q})-2 E_2(\tilde{q}^2)\bigr)\hfill\\
    \hfill+\frac{L^2}{1152}(\ln \tilde{q})^2\left[4 E_4(\tilde{q}^2)+(E_2(\tilde{q}))^2-E_4(\tilde{q})-4(E_2(\tilde{q}^2))^2\right]\\
    =\frac{\pi^4R^2}{1152}\left[4(E_2(q))^2+E_4(\sqrt{q})-(E_2(\sqrt{q}))^2-4 E_4(q)\right]~.\hfill\\
\end{multlined}
\end{equation}
Using equation \eqref{eq_app_sum 8}, we get the integral
\begin{equation}
\label{eq_app_int 8}
\mathtoolsset{multlined-width=0.9\displaywidth}
\begin{multlined}
    \int d \tau_1 d \tau_2  \big(\tilde{G}(\tau_{12})\bigl(\partial_{\tau_1} \partial_{\tau_2} \tilde{G}(\tau_{12})\bigr)^2\big)\Big|_{\sigma_1=\sigma_2=0}\hfill\\
    \hfill=\frac{\pi\ln q}{576L^2}\left[(E_2(\sqrt{q}))^2-4(E_2(q))^2-E_4(\sqrt{q})+4 E_4(q)\right]~.
\end{multlined}
\end{equation}

\bibliography{ref}
\bibliographystyle{JHEP.bst}

\end{document}